# Combinatorial Large-area MoS$_2$/Anatase-TiO$_2$ interface: A Pathway to Emergent Optical and Opto-electronic Functionalities

*Tuhin Kumar Maji [a], J R Aswin [b], Subhrajit Mukherjee [c], Rajath Alexander [d], Anirban Mondal [b], Sarthak Das [e], R. K. Sharma [f], N. K. Chakraborty [g], K. Dasgupta [d], Anjanashree M R Sharma [h], Ranjit Hawalder [i], Manjiri Pandey [j], Akshay Naik [h], Kausik Majumdar [e], Samir Kumar Pal [a], K V Adarsh [b], Samit Kumar Ray [a, c], Debjani Karmakar [k],\**

[a] *Department of Chemical Biological and Macromolecular Sciences, S.N.Bose National Centre for Basic Sciences, Sector III, JD Block, Kolkata 700106, India.*
[b] *Department of Physics, Indian Institute of Science Education and Research, Bhopal 462066, India.*
[c] *Department of Physics, IIT Kharagpur, Kharagpur, West Bengal 721302, India.*
[d] *Advanced Carbon Materials Section, Bhabha Atomic Research Centre, Trombay, Mumbai 400085, India.*
[e] *Department of Electrical Communication Engineering, Indian Institute of Science, Bangalore 560012, India.*
[f] *Raja Rammana Centre for Advance Technology, Parmanu Nagar, Sahkar Nagar Extension, 1, CAT Rd, Rajendra Nagar, Indore, Madhya Pradesh 45201.*
[g] *Material Science Division, Bhabha Atomic Research Centre, Trombay, Mumbai 400085, India.*
[h] *Centre for Nano Science and Engineering, Indian Institute of Science, Bangalore, Karnataka, 560012.*
[i] *Centre for Materials for Electronics Technology, Off Pashan Road, Panchwati, Pune- 411008, India.*
[j] *Accelerator Control Division, Bhabha Atomic Research Centre, Trombay, Mumbai 400085, India.*
[k] *Technical Physics Division, Bhabha Atomic Research Centre, Trombay 400085, India.*

**\*Corresponding Author: Dr. Debjani Karmakar**
E-mail: ***debjan@barc.gov.in***

# Abstract

Interface of transition metal dichalcogenide (TMDC) and high-$k$ dielectric transition metal oxides (TMO) had triggerred umpteen discourses due to the indubitable impact of TMO in reducing the contact resistances and restraining the Fermi-level pinning for the metal-TMDC contacts. In the present work, we focus on the unresolved tumults of large-area TMDC/TMO interfaces, grown by adopting different techniques. Here, on a pulsed laser deposited (PLD) $MoS_2$ thin film, a layer of $TiO_2$ is grown by using both atomic layer deposition (ALD) and PLD. These two different techniques emanate $TiO_2$ layers with different crystalline properties, thicknesses and interfacial morphologies, subsequently influencing the electronic and optical properties of the interfaces. In contrast to the earlier reports of $n$-type doping for exfoliated $MoS_2/TiO_2$ interfaces, large-area $MoS_2$/Anatase-$TiO_2$ films had demonstrated a $p$-type doping of the underneath $MoS_2$, irrespective of the adopted deposition technique. In addition, they manifest a boost in the extent of $p$-type doping with increasing thickness of $TiO_2$, as emerged after analyzing the core-level shifts of the X-ray photoelectron spectra (XPS). Density functional analysis of the $MoS_2$/Anatase-$TiO_2$ interfaces, for pristine and in presence of a wide range of interfacial defects, could explain the interdependence of doping and the terminating atomic-surface of $TiO_2$ on $MoS_2$. The theoretical results resemble well with the realistic scenario of large-area growths after incorporating minimization of surface-strain via mutual rotation of the constituent layers. The optical properties of the interface, encompassing the photoluminescence (PL), transient absorption and $z$-scan two-photon absorption indicate the presence of defect-induced localized mid-gap levels in $MoS_2/TiO_2$ (PLD), resulting quenched exciton signals. On the contrary, the relatively defect-free interface in $MoS_2/TiO_2$ (ALD) demonstrates a clear presence of both A and B excitons of $MoS_2$. This outcome corroborates with the first-principles observation of the presence of localized traps at the interfacial band-structure in presence of the point defects. From the investigation of optical properties, we indicate that $MoS_2/TiO_2$ (PLD) interface may act as a promising saturable absorber, having a significant non-linear response for the sub-bandgap excitations. Moreover, $MoS_2/TiO_2$ (PLD) interface had resulted a better photo-transport. A potential application of $MoS_2/TiO_2$ (PLD) is demonstrated by the fabrication of a $p$-type photo-transistor with the ionic-gel top gate. This endeavor to analyse and understand the $MoS_2/TiO_2$ interface establishes the prospectives of large-area interfaces in the field of optics and optoelectronics.

# 1.Introduction

Recent ages have discerned a surge of efforts for imprinting the superior applicability of two-dimensional (2D) materials in the field of ultra-scaled logic devices over the conventional semiconductor industry materials Si or Ge. Universal set of 2D materials, embracing numerous subsets like Graphene [1-2], Silicene [3], Germanene [4], Phosphorene [5] *etc.* have already manifested their better performances owing to their intrinsic physical attributes like dangling bond free surface containing natural passivation [6], tunable optimal band-gap[7], advanced carrier control [8-9] and the possibility of modulating electro-optical properties by forming combinatorial van der Waal stacks[10-11]. Electronic device utilization of TMDC systems like $MoS_2$ poses a multidisciplinary challenge to the scientific community after demanding a thorough understanding of surface and interfacial chemistry, physics of band alignment along with its consequential impact on charge transport and contact-engineering at the metal-semiconductor interfaces[12-13].

To attain an optimized mechanism of carrier injection and channel transport in TMDC devices, several pathways have been adopted, *viz.* 1) tuning the contact resistances at the metal-semiconductor junction to control the metal-induced gap states (MIGS) and Fermi-level Pinning (FLP)[13-15], 2) dielectric screening of the scattering centers resulting from the interfacial carrier traps by using a high-*k* dielectric, ionic gate or 2D insulators[16-18] and 3) modulation of carrier density and passivation of the defect levels by substitutional doping, functionalization with chemical agents or ionic plasma immersion [19-22].

At metal-semiconductor junctions, the problem of high contact resistances significantly hinders their high performance, especially in the case of geometrical scaling down of device sizes. At such junctions, the Schottky barrier height (SBH = $\phi_B$) as deduced from the Schottky-Mott rule, $\phi_B = \phi_m - \chi_{sc}$, possesses a linear relationship with the metal work-function $\phi_m$ and electron affinity $\chi_{sc}$ of the semiconductor[23]. In practical situations, as a result of the charge redistribution across the interface, the presence of the interfacial dipoles causes a significant modification of the band-bending across the interface. Thus, the modified barrier height becomes dependent on the net charge neutrality level $\phi_{CNL}$ of the semiconductor surface and the FLP factor S by the relation, $\phi_B = S(\phi_m - \chi_{sc}) + (1 - S)\phi_{CNL}$. This relation implies that for highly pinned systems ($S = 0$), $\phi_B$ is always fixed at $\phi_{CNL}$[24]. The FLP factor $S = \partial\phi_B / \partial\phi_m$ has an explicit relation with the interfacial density of states $D_s = (1 - S)\varepsilon_i\varepsilon_0 / Sq^2\delta_i^2$, which could be modulated by the optimal thickness ($\delta_i$) of an appropriate dielectric layer with dielectric constant $\varepsilon_i$. For conventional semiconductor like Ge [25-26] or Si[27], insertion of an ultrathin interlayer of high-$k$ oxides between the metal and the semiconductor is considered to be the most promising expedient to assuage the FLP. This oxide layer results into a smaller SBH and lower contact resistances by dwindling both the virtual MIGS and the real interface states [28]. $TiO_2$ is well-accomplished to be a suitable candidate for the interlayer, producing a significant modification of FLP for different surface geometries [23, 29].

The S-factor for $MoS_2$ is ~ 0.16 - 0.35 [6, 15, 30], which is close to the conventional semiconductor like Ge. $TiO_2$, while used as an interfacial layer, is seen to be effective in reducing the SBH and the contact resistances at various metal-$MoS_2$ junctions [23, 31]. The presence of $TiO_2$ on $MoS_2$ also has a tremendous impact on carrier mobility and scattering effects [29]. In realistic interfaces, Coulomb and point defects as well as the surface roughness act as scattering centers, which significantly influence the carrier mobility. Ideally, mobility should only be limited by phonon scattering. The presence of $TiO_2$ helps to screen the Coulomb scattering from charged impurities, reduces surface roughness and also introduces additional remote phonon scattering [32]. All of these impacts act positively toward increasing mobility. An intermediate $TiO_2$ layer is also seen to be helpful in area-dependent carrier transfer processes and in improving the device scalability [33-34]. Unlike the non-scalable means like molecular and substitutional doping or passivation of defect sites, emanating non-degenerate doping[35] and a low on/off ratio, the interface engineering of $TiO_2$ has proved to be effective to obtain a significant improvement of photodetector/phototransistor performance [36-38]. In addition to $TiO_2$, various other oxides like $MoO_3$, $WO_3$, MgO or self-limiting surface oxides are seen to provide the surface charge transfer doping (SCTD) and thus modulate the carrier density from $n$ to $p$-type in $MoS_2$ [39-41] $WSe_2$[42], $MoTe_2$ [43] and InSe [44-45] respectively. 2D insulators like hexagonal Boron Nitride (hBN) has also been utilized as an encapsulation of the TMDC layer. Thereby it provides a tunneling barrier to modify the work function of the metals and helps to obtain low-temperature ohmic contacts for TMDC [46-47]. The $MOS_2/TiO_2$ interfaces are also commonly used for the betterment of photo-electrochemical and photo-catalytic performances of $MoS_2$[48-49].

Thus, the alluring interface $MoS_2/TiO_2$ has already been thoroughly investigated from the point of view of nanoelectronic and photocatalytic device applications. However, most of the abovementioned studies are done either for micron-sized exfoliated flakes or for small-area CVD growths of $MoS_2$, on which, amorphous ultrathin $TiO_2$ layer was deposited by atomic layer deposition (ALD). For large-area growths of both $MoS_2$ and Anatase $TiO_2$, the interface is yet to be scrutinized thoroughly. Moreover, there are very few studies to analyze and compare the modulation of carrier density and the interfacial impacts of the $TiO_2$ layer with different crystallinity on the optical and optoelectronic properties of the large-area $MoS_2$/A-$TiO_2$ interfaces. These lacunae in the former literature provides the motivation for the current investigation.

The present work is organized as follows. First, crystalline thin films of $MoS_2$ were deposited on a quartz substrate by using PLD. As an overlayer, $TiO_2$ was deposited by using two different techniques, *viz.* PLD of anatase-$TiO_2$ (A-$TiO_2$) and ALD of amorphous $TiO_2$. These two techniques lead to the formation of interfaces with different thicknesses and morphologies. Next, initial structural characterizations are done by using micro-Raman (MR) and cross-sectional transmission electron micrograph (TEM). The binding energy shifts of the core-level spectra of Mo-$3d$ and S-$2p$ for both $MoS_2/TiO_2$ (PLD) and $MoS_2/TiO_2$ (ALD) have been investigated and analysed to obtain an indication of the type of doping of the underneath $MoS_2$. The experimental observation of doping was validated by the first-principles investigation of the $MoS_2$/A-$TiO_2$ interfaces, indicating that the type of doping in such kind of TMDC/TMO hetero-structure has a strong dependence on the surface termination of the TMO layer. The effects of defects and acceptor levels on the properties of the hetero-structures are also emphasized. As a next step, we have compared the optical properties of both of these heterostructures by room temperature and low-temperature photoluminescence (PL). The dynamical optical properties of the excited states are explored by measuring the transient absorption spectra and the non-linear responses are compared by the *z*-scan two-photon absorption measurements. Finally, we measured the four-probe photo-transport for both of these systems and then demonstrated a top-gate photo-transistor from the $MoS_2/TiO_2$ (PLD) system. At the end, we summarize our results with a conclusion.

## 2. Experimental and Theoretical Details:

### 2.1. Experimental Details:

Commercially available $MoS_2$ (99.99% purity) and Anatase $TiO_2$ (99.9%) powders were finely ground and the resulting powder was compressed to form a pellet of one-inch diameter and ~ 5 mm thickness at a pressure of 2 Mpa in a uni-axial press. The pellets are then vacuum sealed in two separate quartz cylinder at a pressure ~ $10^{-6}$ mbar. The cylinders containing $MoS_2$ and $TiO_2$ are then sintered at $800^oC$ and $1000^oC$ respectively for 5 hours. After sintering, these pellets are structurally cross-checked to resemble the original phase.

The pellets were mounted on both of the target holders of the PLD chamber. The chamber was evacuated by a combination of turbo molecular and rotary pump to a base pressure of ~$1.33\times10^{-6}$ mbar. Clean Quartz substrates of size 10mm×10mm×1mm were mounted on the sample holder. KrF excimer laser with a wavelength of 248 nm was used for pulsed deposition. The substrate annealing temperature is kept at $650^oC$. For the deposition of $MoS_2$, a repetition rate of 5 Hz and annealing time of five minutes were used. $TiO_2$ was deposited with a repetition rate of 10 Hz with the same annealing time.

In the second set of hetero-structures, $TiO_2$ thin films were deposited on $MoS_2$ by Atomic layer deposition (ALD) with Titanium (IV) isopropoxide (TTIP) in 25 g packages within a stainless steel cylinder. Deionized water ($H_2O$) was used as the second precursor as well. The pressure of the ALD chamber was maintained at 50 kPa, and the flow-rate of the $N_2$ carrier gas was set at 20 sscm. The TTIP was heated to 250°C and the water was kept at room temperature in order to provide sufficient vapor for the ALD process. The precursor pulse rates for TTIP and $H_2O$ were set at 0.20 s and 0.15 s respectively with the ALD cycles sequence $H_2O$ / $N_2$ / TTIP/ $N_2$. The deposition temperature is kept at 250°C and a blanket deposition of 2 nm thickness was performed with a growth per cycle 0.04 Å/second.

High-resolution cross-sectional transmission electron micrographs are obtained with a JEOL make field-emission gun transmission electron microscope (model JEM 2100F) of magnification 50X – 1.5MX and accelerating potential of 200 KV.

The X-ray photo-electron spectroscopic (XPS) measurements were done at BL-14 Beamline of RRCAT – Indore, which is used for high-resolution photoelectron spectroscopy of solids at hard X-ray energy-range (2-15 keV). In this measurement, the HS are bombarded with high energy X-ray and the resulting energy spectrum of emitted photo-electrons is analyzed. This spectrum is used to calculate the binding energy of the core electrons in the material, providing the elemental information and the valence configuration of the exposed surface. Whereas, typical lab XPS provides information up to a depth of about 1 nm, the high energy XPS is capable of extracting information at larger depths, as the mean free path increases with electron energy. The synchrotron source provides high-intensity X-rays over a wide range of energies and XPS at different energies provides chemical information as a function of depth. The beamline consists of a Pt-coated toroidal mirror for focusing the X-ray beam, a double crystal monochromator for wide tunable X-Ray energy range up to 15 keV, water cooled X-ray slits for defining beam opening and high-resolution hemispherical analyzer based experimental station for acquiring the XPS data.

In ultrafast exciton dynamics measurements, 120 fs pulses, having their center at 800 nm and with a repetition rate of 1 kHz, were divided into two beams, designated as pump and probe. The former beam is passed through an optical parametric amplifier and is considered as the pump beam (400 nm, 120 fs pulses and a fluence of 300 μJ/cm$^2$). The latter beam is a white light continuum delayed with a computer-controlled motion controller with respect to the pump. The pump and probe beams were spatially overlapped on the sample, and the change in absorbance of the probe beam $\Delta A = -(log[I_{ex}/I_0])$ was determined, where $I_{ex}$ and $I_0$ are the transmitted intensities of the sequential probe pulses after a delay time $\tau$ following excitation by the pump beam, and in the ground (in dark) state, respectively.

An open aperture Z scan is used to obtain a quantitative analysis of the third-order nonlinearity in terms of the analysis of nonlinear refractive index and nonlinear absorption coefficient. In this measurement, two photodiodes are used as detectors, with one collecting the input laser energy (D1) and the other measuring the transmitted energy through the thin film (D2). By calculating the ratio of D2/D1, we measure the normalized transmittance. Here normalized transmittance is measured as a function of the sample position. It is a moving system through a translational stage with varying intensity profile throughout the measurement. Z scan features are symmetric in nature because of the intensity profile prepared by a convex lens, where the maximum intensity is observed at the focal point. In nanosecond open aperture Z scan, Nd: YAG laser of 1064 nm with pulse-width of 7 ns was used to excite the thin film under investigation. A fixed repetition rate of 10 Hz is used to avoid the damage of the thin film via the heating effect. The beam waist and the Rayleigh length in our experiment were 19 μm and 1.6 mm, respectively. The Gaussian beam is focused with the help of a 20 cm plano-convex lens of the *z*-axis of the computer-controlled translation stage.

**2.2. Computational Details:**

For first-principles analysis, spin-polarized plane-wave pseudopotential calculations were performed with norm-conserving projector augmented wave (PAW) pseudopotentials as implemented in the Vienna Ab-initio Simulation Package (VASP)[50]. The valence levels for Mo consist of 4*p*, 5*s,* and 4*d* orbitals and those for S constitute 3*s* and 3*p* orbitals. The exchange-correlation interactions are incorporated with generalized gradient approximation (GGA) with Perdew-Burke-Ernzerhof (PBE) functionals after taking into account the spin-orbit coupling (SOC). Interface-induced dipolar interactions are integrated after taking care of van der Waals corrections after including a semi-empirical dispersion potential to the DFT energy functional following the Grimme DFT-D2 method[51]. The cut-off energy for the plane wave expansion is set as 500 eV and a Monkhorst-Pack grid 5×5×3 is used for Brillouin zone sampling for all calculations. The ionic positions and the lattice parameters are relaxed by using the conjugate gradient algorithm until the Hellmann-Feynman force on each ion is less than 0.01 eV.

# 3.Results and discussion

## 3.1. Fabrication of the Heterostructure

We have employed the PLD technique under ultra-high vacuum with the vacuum-sintered target pellets of 2H-$MoS_2$ and A-$TiO_2$ respectively for deposition of pristine $MoS_2$ and $MoS_2/TiO_2$(PLD). The thickness of the $MoS_2$ film ranges from 25-28 nm, equivalent to more than 30 layers of $MoS_2$. The number of pulses used for $MoS_2$ deposition are 400, 600, and 1000. To avoid the discrete deposition and to obtain a continuous overlayer of $TiO_2$, the number of pulses needed for the growth of $TiO_2$ are optimized to 3600, which leads to a deposition of thickness ~ 80 nm. The quartz substrate is having a dimension of 10 mm × 10 mm × 0.5 mm. Another interface $MoS_2/TiO_2$ (ALD) is grown after using the cyclic gas-surface self-limiting reactions in the ALD setup for a uniform deposition of 2 nm. The details of the deposition techniques are described in the Supporting Information. In both of these cases, the surface roughness of $MoS_2$ decreases by an order of magnitude after the deposition of TiO$_2$. In Supporting Information, the atomic force microscopy measurement of the thickness and an estimate of roughness for deposited thin films are described (Fig S1). While the ALD TiO$_2$ layer is known to be amorphous [23, 38], the PLD TiO$_2$ layer is seen to be polycrystalline. The details of the thicknesses and the surface roughnesses of various systems are tabulated in Table S1.

Throughout the study, we will follow the convention of designating the MoS$_2$ (PLD)/TiO$_2$ (ALD) and MoS$_2$ (PLD)/TiO$_2$ (PLD) hetero-structures (HS) as HS1 and HS2 respectively. Moreover, HS1 (1000), HS1 (600), and HS1 (400) will imply the HS1 with MoS$_2$ deposited by using 1000 pulses and so on. In a similar way, MoS$_2$ (1000) will indicate the MoS$_2$ PLD films deposited by using 1000 pulses.

### 3.2. Structural Characterization: Micro-Raman and Transmission electron micrograph

Fig 1 (a) – (c) depicts the micro-Raman spectra of pristine MoS$_2$ (400, 600, 1000 pulses) and the corresponding HS1 and HS2 systems upon illumination with a 532 nm green laser light. To verify the uniformity of the respective films over a large area, the spectra was produced by taking the average of several Raman mappings over a large area of 400 μm × 400 μm at various places of each film. The inset in the same figure shows the microscopic image of the corresponding thin-film representing their continuous large-scale nature. Fig 1(a) depicts the characteristic peaks for the in-plane Mo–S vibration ($E^1_{2g}$) and out-of-plane S-vibration ($A_{1g}$) with the enhanced intensity for the later one for pristine as well as for both the HS. Such enhancement occurs due to the strong electron-phonon coupling for resonant Raman scattering at the K-point, where the electrons at $c$ axis-aligned Mo-$d_{z2}$ orbital are coupled with the phonons along the same direction due to the atomic displacements responsible for $A_{1g}$ [52]. The peak positions for both of the characteristic peaks are listed in Table S2, indicating that with an increasing number of the pulses, there is an increase in the film-thickness of MoS$_2$, as can be seen from the increasing difference between the $E^1_{2g}$ and $A_{1g}$ peak positions [53]. With the deposition of the amorphous TiO$_2$ by ALD, the MoS$_2$ peak positions experience a blue-shift, similar to the earlier study [38], whereas, with polycrystalline TiO$_2$ in PLD, there is an overall red-shift, indicating the detrimental effect of the interfacial strain on both the in-plane and out-of-plane vibrations for the thicker TiO$_2$ layer. The crystalline nature of the film may have undergone a slight decrement, as indicated by a slight increase of the full width at half maxima (FWHM) after the deposition of TiO$_2$ for both HS1 and HS2. For the hetero-structure HS1, the signature of TiO$_2$ is not visible in the Raman spectra (Fig 1(b)), as is expected for an amorphous TiO$_2$ film [54]. On the contrary, HS1 is having the signature peak of the anatase phase of TiO$_2$ at ~150 $cm^{-1}$ as a result of the better crystallinity of the TiO$_2$ layer [54].

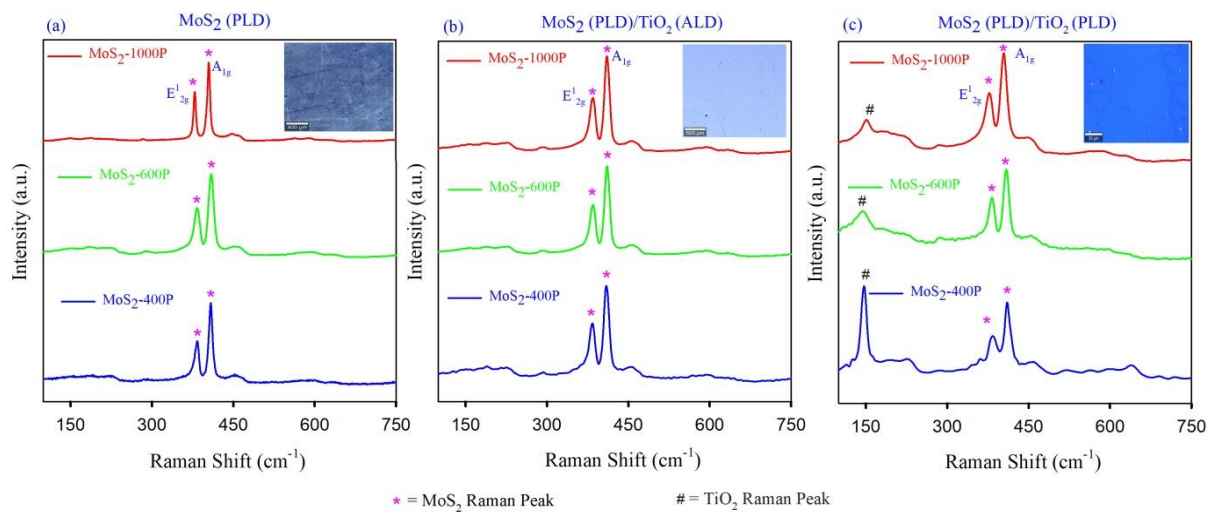

*Figure 1:* a) Micro-Raman spectra of a) MoS$_2$ (PLD), b) MoS$_2$ (PLD)/TiO$_2$ (ALD) and c) MoS$_2$ (PLD)/TiO$_2$ (PLD) for different thicknesses of MoS2 (400, 600, 1000 pulses) and same thickness of TiO$_2$ (3600 pulses) for PLD samples.

Figure 2 depicts the cross-sectional high-resolution transmission electron microscopic (HRTEM) images for both MoS$_2$ (1000) and HS2 (1000) films on quartz. Fig 2(a) clearly shows the layered nature of MoS$_2$ on quartz, where, the lattice spacing, as calculated from the magnified fringe pattern of Fig 2(b), is 6.55Å, indicating a stacking along the [0001] direction [48]. Fig 2(c) presents the cross-sectional image of HS2 films on quartz, with different layers indicated. For the same sample, the magnified image near the MoS$_2$/TiO$_2$ interface (Fig 2(d)) indicates the crystalline and polycrystalline nature of the PLD-grown MoS$_2$ and A-TiO$_2$ respectively. The random orientations of the multiple grains are visible from the magnified image of the A-TiO$_2$ part of the interface in Fig 2(e), where the fringe-width of 2.5Å matches with the [103] plane of A-TiO$_2$. Fig 2(d) also manifests an uneven interface indicating the presence of defects like vacancies or edge boundaries. Ideally, the structure of MoS$_2$ should be devoid of any dangling bonds and carrier trapping defect sites. However, realistic HRTEM image of MoS$_2$ PLD layer indicates the presence of discontinuous edges. These edges can be passivated by the TiO$_2$ layer, improving the overall crystallinity of the combined hetero-structure. This will also be evident from the selected area electron diffraction (SAED) images, containing both spots and ring pattern, as seen in Fig

2(f) and also resembling with the prior studies [34, 38].

Thus, after fabrication of both of the proposed HS, the structural characterizations by using the Micro-Raman and TEM validate the synthesis of the HS.

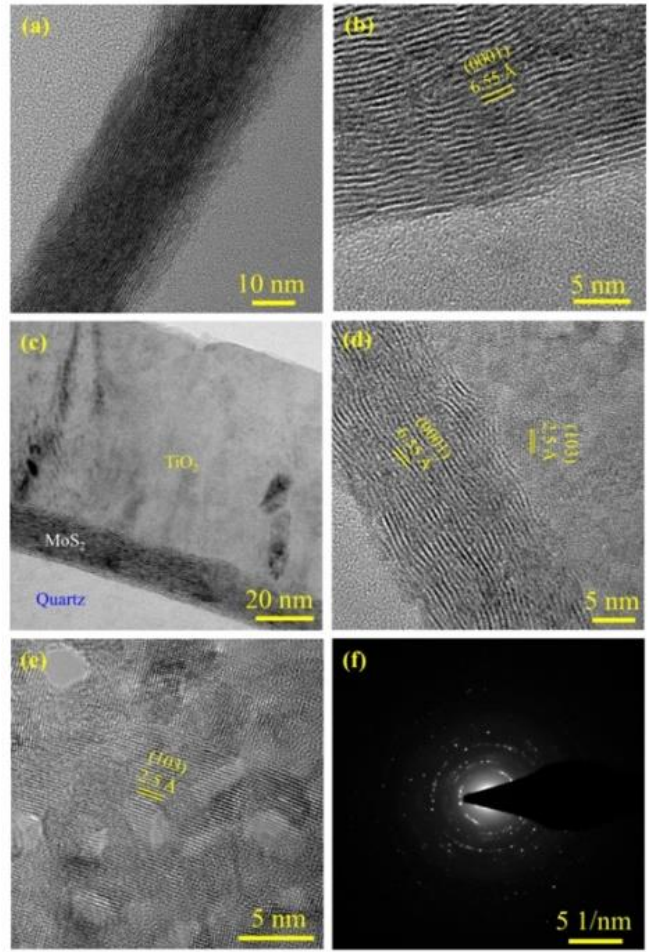

*Figure 2:* a) Cross-sectional TEM image of PLD-grown $MoS_2(1000)$, b) HRTEM image of $MoS_2(1000)$, c) cross-sectional micrograph of HS2(1000) (see text), showing Quartz, $MoS_2$, and $TiO_2$, d) HRTEM image of HS2(1000), e) HRTEM image of $TiO_2$ layer of HS2(1000) and f) SAED image of HS2(1000).

**3.3. X-ray photoelectron spectroscopy: BE shifts of core level spectra**

The standard identification technique adopted to find out the nature of doping in 2D HS is the measurement of the BE shifts for the core level electrons in the XPS spectra [21, 23, 55-56]. To obtain an idea of the elemental composition and the corresponding valence configuration, high-resolution photoelectron spectroscopy of the films are performed with hard X-rays in an energy-range of 2-15 keV. In a similar way like the Raman spectra, the resultant XPS data for each film is obtained by averaging over several scans at various places of the respective film. The details of the measurement are presented at the Supporting Information.

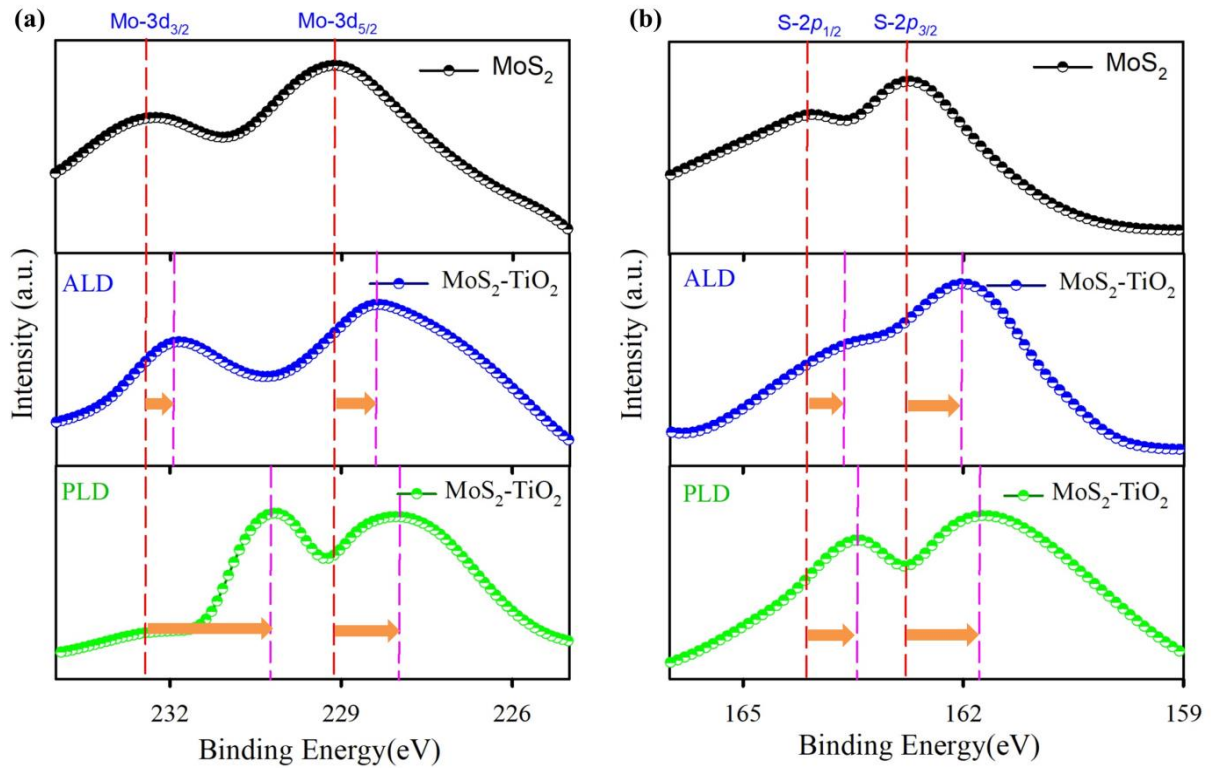

***Figure 3:*** *XPS spectra depicting the core level shifts for MoS$_2$ (1000P)/TiO$_2$ systems: a) comparison of Mo-3d levels for pristine MoS$_2$, HS1 (see text) and HS2 (see text); b) same comparison for S-2p levels.*

Fig 3 depicts the comparative plots for the binding energy (BE) of the core levels, *viz.*, Mo $3d_{5/2}$, Mo $3d_{3/2}$, S $2p_{3/2}$ and S $2p_{1/2}$ for the pristine MoS$_2$ and for both HS1 and HS2 with MoS$_2$ deposited with 1000 pulses. After deposition of TiO$_2$, both the Mo-3*d* (Fig 3(a)) and S-2*p* (Fig 3(b)) doublets undergo a shift towards the lower BE for HS1 and HS2, implying a shift of the Fermi-level towards the valence band and thereby indicating a *p*-type doping. As is also evident from the Fig 3, the shift of the BE and thus the extent of the *p*-type doping is more for HS2 (PLD-TiO$_2$) than HS1(ALD-TiO$_2$). The shift of different core levels are observed to be different, which can be attributed to the non-uniformity of the MoS$_2$/TiO$_2$ interfaces, whereby the environments around the Mo and S atomic levels will differ. Supporting Information contains the comparative XPS spectra of HS1 and HS2 with the corresponding pristine system deposited with 400 and 600 pulses of MoS$_2$. For all of these HS, the shifts of BE indicate a *p*-type doping, which differs from the prior results in the literature for the exfoliated MoS$_2$ and ALD TiO$_2$ [23, 34, 38]. The consolidated list of shifts in BE for different HS is presented in Table S3.

Therefore, for both HS1 and HS2, the large-area HS are resulting into a *p*-type doping of the

MoS$_2$ layer, contrary to the prior studies on exfoliated HS. As a next step, we analyse this doping behaviour with the help of First principles calculations.

**3.4. First principles study of interface: correlation of surface termination and doping**

To understand the doping behaviour of the large-area interfaces, we have elaborately investigated the MoS$_2$/TiO$_2$ interface. There is a significant difference between the interface created from the micron-sized exfoliated MoS$_2$ and a fast deposition of amorphous TiO$_2$ by ALD [21, 23, 34, 38] with the large-area epitaxial growth processes used in the present study. While the large-area interfaces are grown with longer deposition times, there are provisions of a gradual relaxation of the interfacial strains with increasing thickness of the top layer by mutual rotation of the constituting surfaces at the interfacial region. Keeping this scenario in mind, construction of the interface is done after stacking a (2×2×1) supercell of 2H-MoS$_2$ (P6$_3$/*mmc*) with a (2×2×1) supercell of A-TiO$_2$ (I4$_1$/*amd*) by the co-incidence site lattice (CSL) method, as implemented in the ATOMISTIC TOOLKIT 15.1 package[57-58]. For minimization of the mutual rotational strain of the interface, a survey was performed through the grid $m\mathbf{v}_1 + n\mathbf{v}_2$, with the vectors $\mathbf{v}_1$ and $\mathbf{v}_2$ being the basis vectors of the MoS$_2$ lattice, so that for the maximum values of the integers $m$ and $n$ (6 in the present case), the supercells of both of the lattices have the lowest mismatch. In the next step, the mutual strain is minimized by varying the relative rotational angle between the MoS$_2$ and TiO$_2$ surface cells around the stacking direction (*c*-axis for the present case) in increments of one degrees. In all of our calculations, the minimized mutual strain is ~0.87%, when the mutual rotational angle between the two surfaces is ~ 46 degrees. The grid areas for MoS$_2$ and TiO$_2$ and their respective relative rotations are presented in Fig S4. Fig 4 represents the two possibilities of surface termination of the TiO$_2$ layer on MoS$_2$, *viz.* a) O-termination and b) Ti-termination.

The electronic structures of these two interfaces are investigated with the help of GGA-PBE + SOC calculations, as implemented in Vienna Ab-initio Simulation Package (VASP)[50]. The details of the methodology for the formation of HS are mentioned in Supporting Information. We have taken care of the impact of the presence of vacancies for these two interfaces, where, for the O-terminated interface, the interfacial S (at MoS$_2$) and O (at TiO$_2$) vacancies are designated as SV and OV respectively. In a similar way, for the Ti-terminated one, the S (at MoS$_2$) and Ti (at TiO$_2$) vacancies are denoted as SV and TiV respectively. The outcome of a converged calculation of the Fermi-level shifts for all these interfaces is presented in

Table 1, revealing the significant impact of surface termination and the interfacial vacancies on the nature of doping. Whereas the Ti-terminated surfaces are heavily *n*-type doped with a significant shift of $E_F$ towards the conduction band (Table 1) revealing a metallic nature, the O-terminated ones indicate a *p*-type doping with a shift of $E_F$ towards the valence band (Table 1). Additionally, in the latter case, the system is semiconducting in one spin channel and metallic in the other one, representing a perfect half-metal. Since the experimental systems are all indicating a *p*-type doping, we present the results of the O-terminated interface in the main manuscript, with the Ti-Terminated details being at Supporting Information.

Fig 5 depicts the layer cum orbital-projected band-structure and the corresponding density of states (DOS) for the pristine O-terminated interface and the same interface with SV and OV respectively. The vertical up and down panels constitute the layer-projected $MoS_2$ and $TiO_2$ fatbands with the corresponding orbital contributions as Mo-*d* (red), S-*p* (green), Ti-*d* (magenta) and O-*p* (blue) respectively. A thorough scrutiny of the band-structures indicates the following key findings:

A) The direct-band gap resulting from Mo-*d* orbitals(red bands in the upper panel of Fig 5(a)) remains unaltered in the HS with a *k*-space shift of the conduction band minima in between the Γ-M high-symmetry directions as a result of the van der Waal interactions with the vertically stacked $TiO_2$. The overall semiconducting nature of $MoS_2$ is protected.
B) $TiO_2$ results into the acceptor levels within ~ 0.5 eV of the valence band of $MoS_2$ (blue bands in the lower panel of Fig 5(a)), as a result of the charge transfer from the mutually rotated $MoS_2$ to $TiO_2$. It also leads to the generation of localized mid-gap states at ~ 1eV. Interestingly, all these $TiO_2$ generated states are contributing only to the up-spin channel and are mostly having the O-*p* orbital character leading to the half-metallic nature of the interface, as can also be seen from the corresponding orbital projected DOS.
C) For $MoS_2$, there is a mixing of the Mo-*d* and S-*p* characters both at valence band maxima (VBM) and conduction band minima (CBM). On the other hand, for $TiO_2$, the Ti-*d* orbital contributions are prominent at CBM with the O-*p* orbitals contributing near VBM.
D) In presence of OV (Fig 5(b)), some additional mid-gap states having O-*p* orbital characters are generated at ~ 1eV only in the up-spin channel, retaining the half-metallic nature. However, in case of SV(Fig 5(c)), mid-gap levels having S-*p* orbital characters are generated in both the spin-channels.

E) In addition to the SOC induced splitting at the VBM, there are similar splitting of levels at the CBM too. The unequal splitting of the bands at both the band-edges may lead to the occurrence of satellite peaks adjacent to the primary excitons, as can be seen in the next section.

The corresponding Ti-terminated layer and the orbital projected band-structures are presented in Fig S5. From the same figure, the heavily *n*-type doped metallic nature of the interface with the near $E_F$ contribution resulting mostly from $TiO_2$ is evident. The electronic properties of prior references having an exfoliated $MoS_2$ and ultrathin amorphous $TiO_2$ resemble with this Ti-terminated structure [23, 34, 38].

Fig 6(a) and (d) illustrate the converged charge and spin-densities for the O-terminated interfaces, where most of the charge and spin-densities are observed to be located around the $TiO_2$ layer. The $MoS_2$ layer is clearly devoid of charge, supporting the *p*-type nature of the interface. Remarkably, the magnetic nature of the system originates solely due to $TiO_2$. In presence of OV (Fig 6(b), 6(e)), and SV (Fig 6(c) and 6(f)), in spite of local reshuffling of charge and spin densities, the *p*-type nature of the system remains unaltered. Fig S6 presents the charge and spin-densities for the Ti-terminated HS, where more overlapping charge and spin-regions between the $MoS_2$ and $TiO_2$ layer upholds the *n*-type nature of doping obtained for the interface.

Thus, the first principles investigations of the $MoS_2/TiO_2$ interface indicates the importance of surface termination and the presence of interfacial vacancies on the electronic properties of the HS.

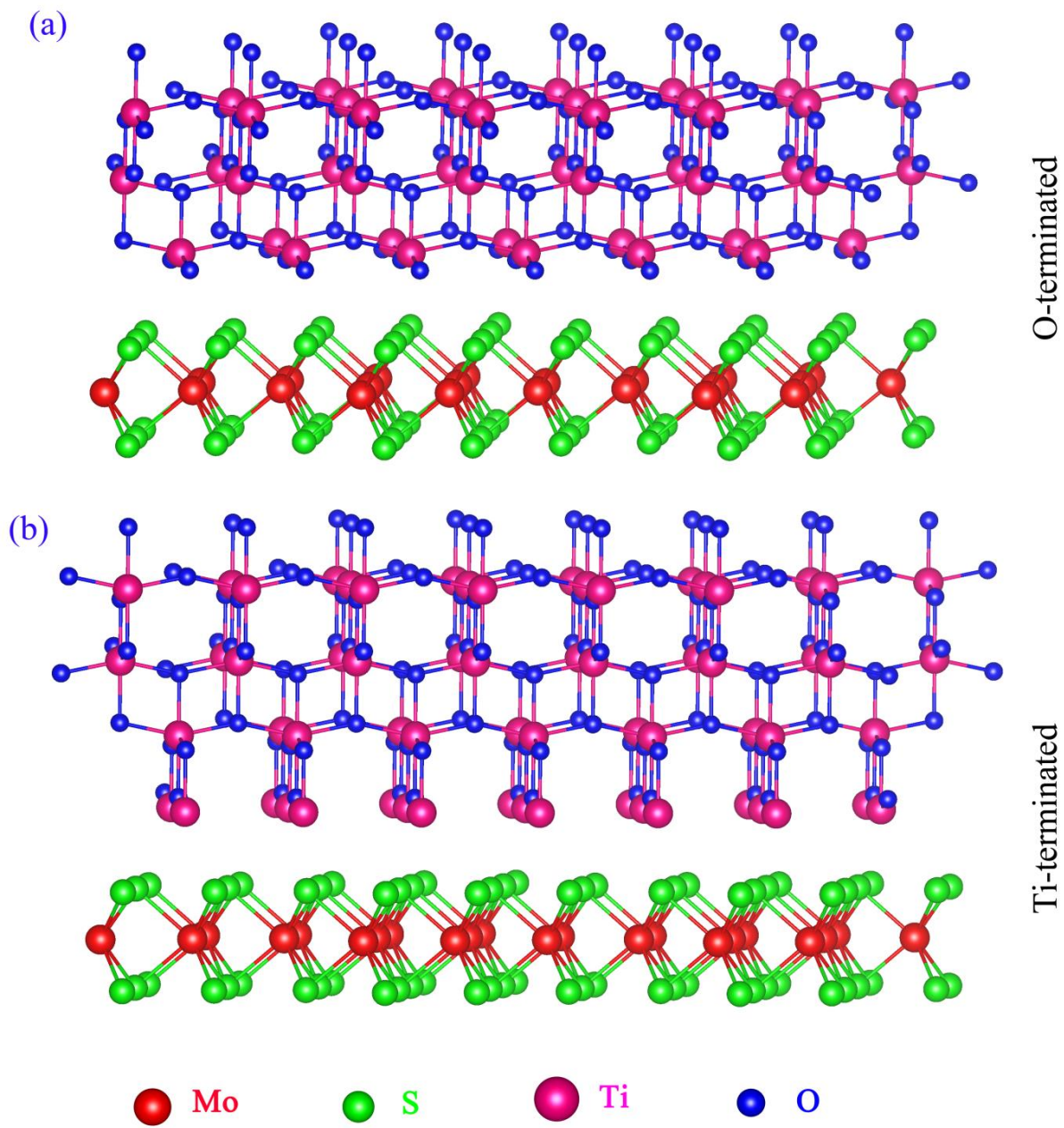

*Figure 4:* *Structural image of a) O-terminated MoS$_2$/TiO$_2$ hetero-structure and b) Ti-terminated MoS$_2$/TiO$_2$ hetero-structure.*

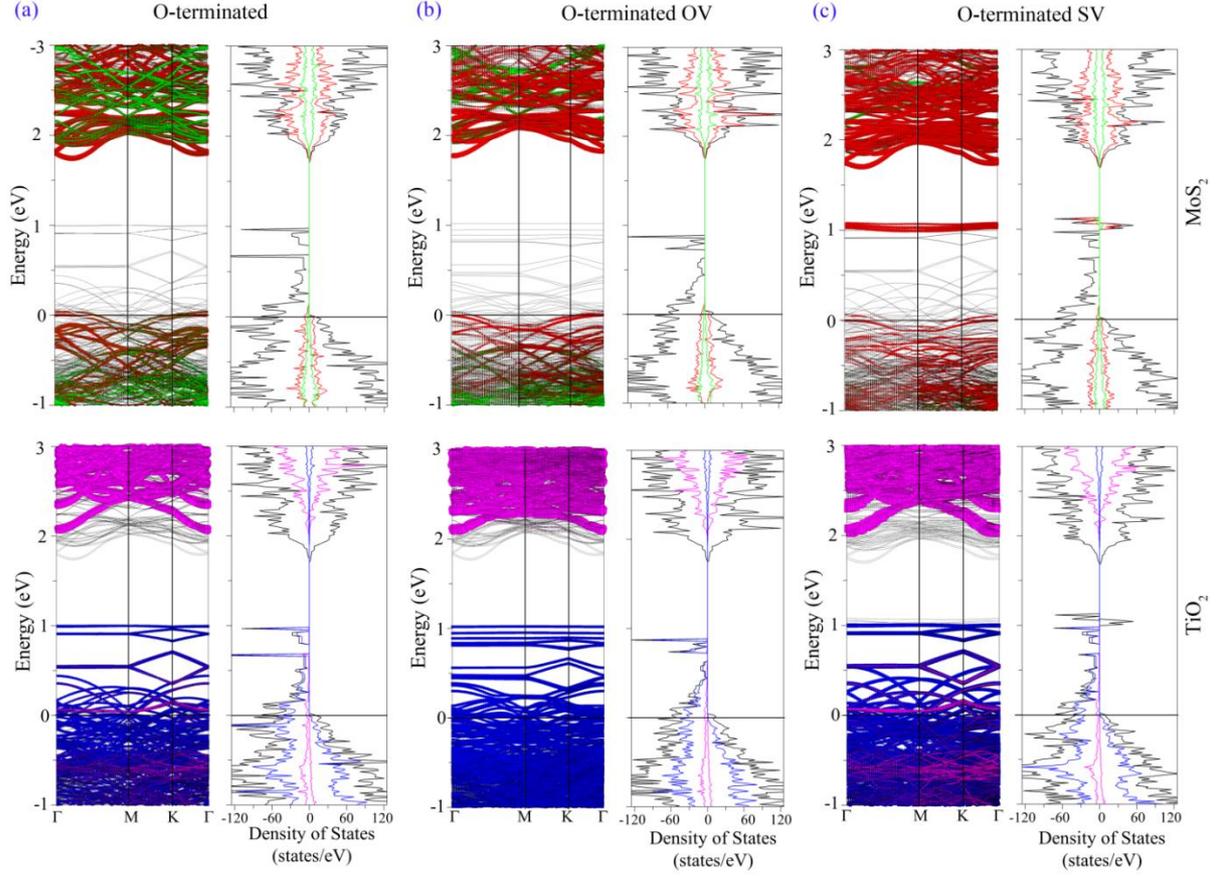

***Figure 5:*** *Layer-projected fatbands for MoS$_2$(upper panel) and TiO$_2$(lower panel) and the corresponding partial density of states (DOS) for a) O-terminated MoS$_2$-TiO$_2$ systems, b) O-terminated MoS$_2$-TiO$_2$ systems with O-vacancy(OV) and c) O-terminated MoS$_2$-TiO$_2$ systems with S-vacancy(SV). The colour code used for orbital projections are Mo-d(red), S-p(green), Ti-d(magenta) and O-p(blue).*

*Table 1: Shift of E$_F$ versus surface termination for different systems (see text).*

| System | Shift of E$_F$ (eV) | Type of doping |
|---|---|---|
| O-terminated | -0.10 | *p*-type |
| O-terminated + SV | -0.11 | *p*-type |
| O-terminated + OV | -0.12 | *p*-type |
| Ti-terminated | +1.53 | *n*-type |
| Ti-terminated + SV | +1.54 | *n*-type |
| Ti-terminated + TiV | +1.72 | *n*-type |

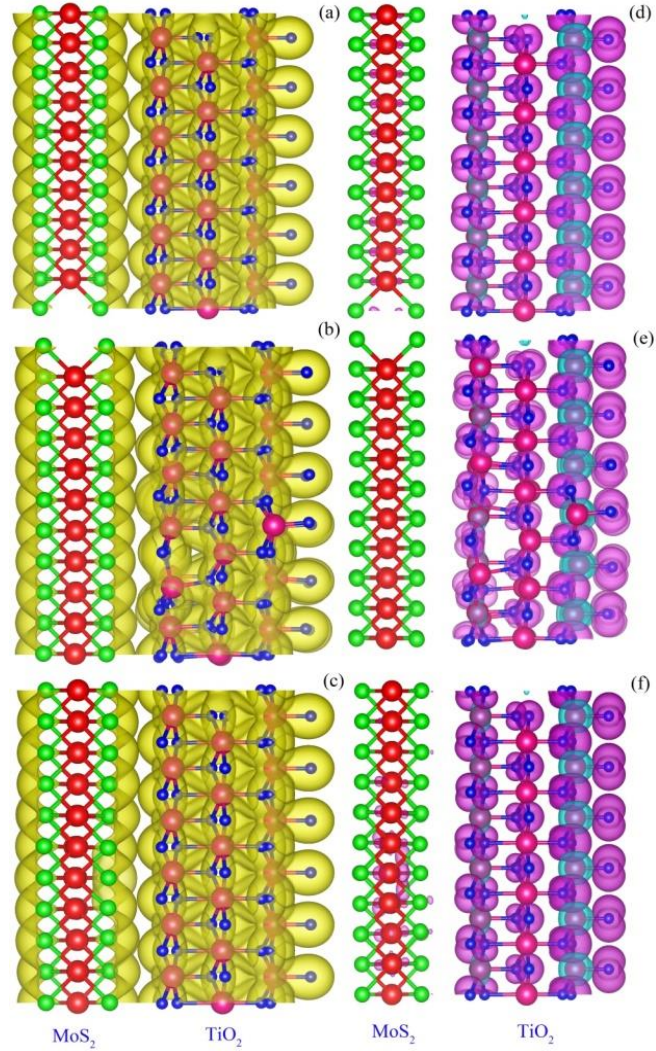

*Figure 6:* *Charge densities of different systems a) O-terminated MoS$_2$/TiO$_2$, b) same with OV, c) same with SV and corresponding spin densities of different systems d) O-terminated MoS$_2$/TiO$_2$, e) same with OV and f) same with SV.*

### 3.5. Photoluminescence of the hetero-structures: impact of TiO$_2$ deposition

Photoluminescence (PL) was measured for the MoS$_2$(1000), HS1(1000) and HS2(1000) with a laser of 532 nm wavelength and spot size of 20 μm diameter with the maximum power of the laser to be restricted to ~ 1.6 mW. In monolayer MoS$_2$, due to the SOC induced VB splitting and the enhanced Coulomb interactions, there are multi-exciton interactions surviving even at room temperature [59-60]. Even for multi-layered systems, there is a symmetry-driven even and odd-layer distribution of inter and intra-layer excitons. The intra-layer excitons, having an order of magnitude higher radiative decay rate, contribute the most to the PL signal. The layer-distribution of the excitons imparts their higher stability even in the bulk limit [61]. Fig 7 depicts the comparative PL spectra for a specific set MoS$_2$ (1000),

HS1 (1000), and HS2 (1000) with the general feature of increased intensity and linewidth for the B exciton. In prior studies on large-area $MoS_2$, there are signatures of the enhanced intensity of B exciton, which are attributed to the proximity of the electron-hole pairs in 2D systems possessing larger densities of excitons, leading to mutually driven correlated inter-excitonic interactions [62]. Fig 7(a) presents the PL spectra of HS1(1000) (red) and pristine $MoS_2$(1000) (green) at 300 K. Pristine $MoS_2$ reveals the A and B exciton peaks at ~ 1.77 and ~ 2.08 eV, revealing a red and blue shift in comparison to the monolayer $MoS_2$ exciton positions respectively. These shifts can be attributed to the multi-layered nature and weak van der Waal stacking due to the presence of defects and surface edges, as is also evident from the TEM images. For HS1 (1000), in addition to the A and B excitons, there is an additional saddle-point peak ~ 0.1 eV away from the B exciton, presence of which may be ascribed to the two direct consequences of the band-structure of the HS. First, the SOC splitting at CBM and second, the presence of $TiO_2$ generated acceptor levels, which may accommodate the de-excited electrons from CBM. Fig 7(b) has a weak A exciton peak at ~ 1.79 eV and a strong B exciton at 2.08 eV. The saddle-point peak is not evident here because of the stronger B exciton. Low temperature (4 K) PL of both HS1 and HS2, however, reveals the prominent presence of all three peaks, *viz*. A exciton (~1.76 eV), saddle point peak (~ 1.96 eV) and B exciton (~ 2.09 eV) as is evident from the Fig 7(c) and (d) with three different laser powers. A comparison with the PL of pristine $MoS_2$ reveals an overall increase of PL intensity in the HS with HS1 (ALD) having a stronger intensity than HS2 (PLD). Supporting Information shows a comparison of PL at 4 K with the laser focussing at different points of the HS (Fig S7).

Therefore, the room-temperature and the low-temperature PL data convey the details of the origin of the excitonic interactions for both the HS. In the next section, we will investigate the exciton dynamics of these HS.

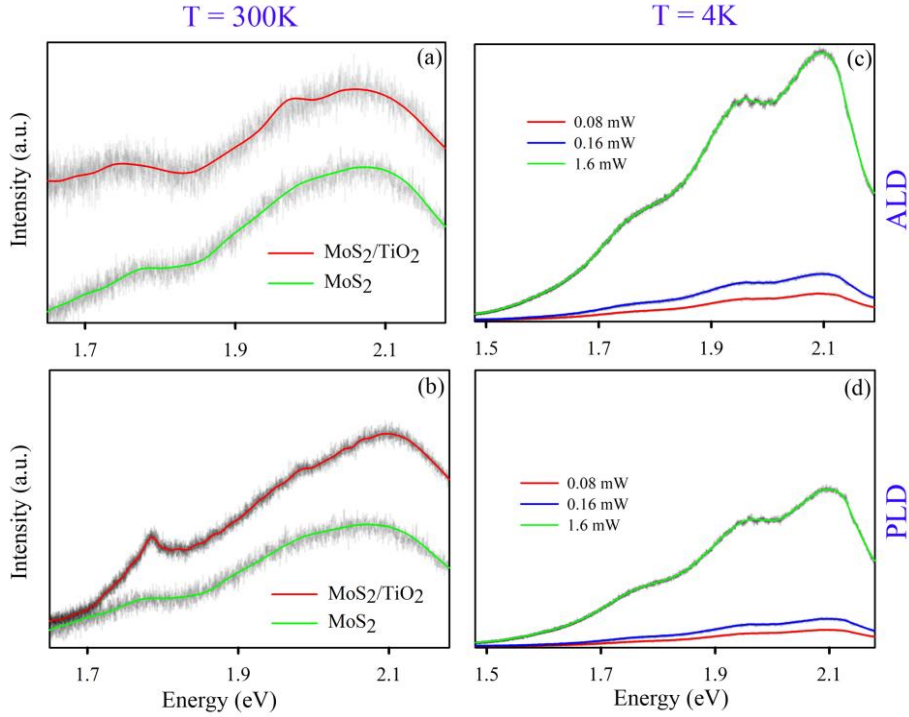

***Figure 7:*** *Comparative PL spectra of a) HS1(1000) and MoS$_2$(1000) at 300 K, b) HS2(1000) and MoS$_2$(1000) at 300 K, c) HS1(1000) at 4 K for three different laser powers and d) HS2(1000) at 4 K for three different laser powers.*

### 3.6. Probing exciton dynamics by transient absorption for hetero-structures

The exciton dynamics of the pristine, as well as the HS, are probed by using ultrafast (picosecond-resolved) transient absorption (TA) spectroscopy, in which, the system is excited with pump photons of 3.1 eV, exceeding the steady state band-gap of MoS$_2$. Next, the changes in the absorbances are measured by using a probe signal ranging from 1.79 to 2.6 eV. Before executing the survey on the excited state, we have measured the ground state absorption spectrum at room temperature, as presented in Fig S8. A closer scrutiny of the overall absorbance in Fig S8, reveals that the quenched nature of the A and B excitons in MoS$_2$ multi-layered films and HS2(1000). HS1(1000) manifests a more prominent A and B exciton features. Fig 8(a)-(d) describes the contour plot of TA for MoS$_2$ (1000), TiO$_2$ (3600), HS1 (1000), and HS2 (1000) respectively, where the pump-probe delay is plotted against the energy of the probe beam. The color codes for the contour, representing the change in the absorbance between the excited and ground states (A$_{excited}$ − A$_{ground}$ = m$\Delta$A where m = 10$^{-3}$), are depicted as side-bar. Figs 8(e)-(g) represents the m$\Delta$A as a function of the pump-probe delay for MoS$_2$ (1000), HS1(1000) and HS2(1000) respectively. The figures show bleach at the exciton position due to the filling up of the excited state and a time delay-induced absorption below the exciton position due to the presence of the trap states. The time-scales for the exciton decay and the trap state filling are represented in table 2. The TA contour and the kinetics plots for MoS$_2$ (400), MoS$_2$ (600) and also for the corresponding HS1 and HS2 are presented at Supporting Information.

*Table 2: Exciton decay and trap build-up times for different systems.*

| Sample | Exciton decay (ps) | Trap build up (ps) |
|---|---|---|
| MoS2 (1000) | 0.29±0.01 ps | 0.33±0.01 ps |
| HS1 (1000) | 0.41±0.01 ps | 0.42±0.01 ps |
| HS2(1000) | 0.19±0.01 ps | 0.22±0.02 ps |

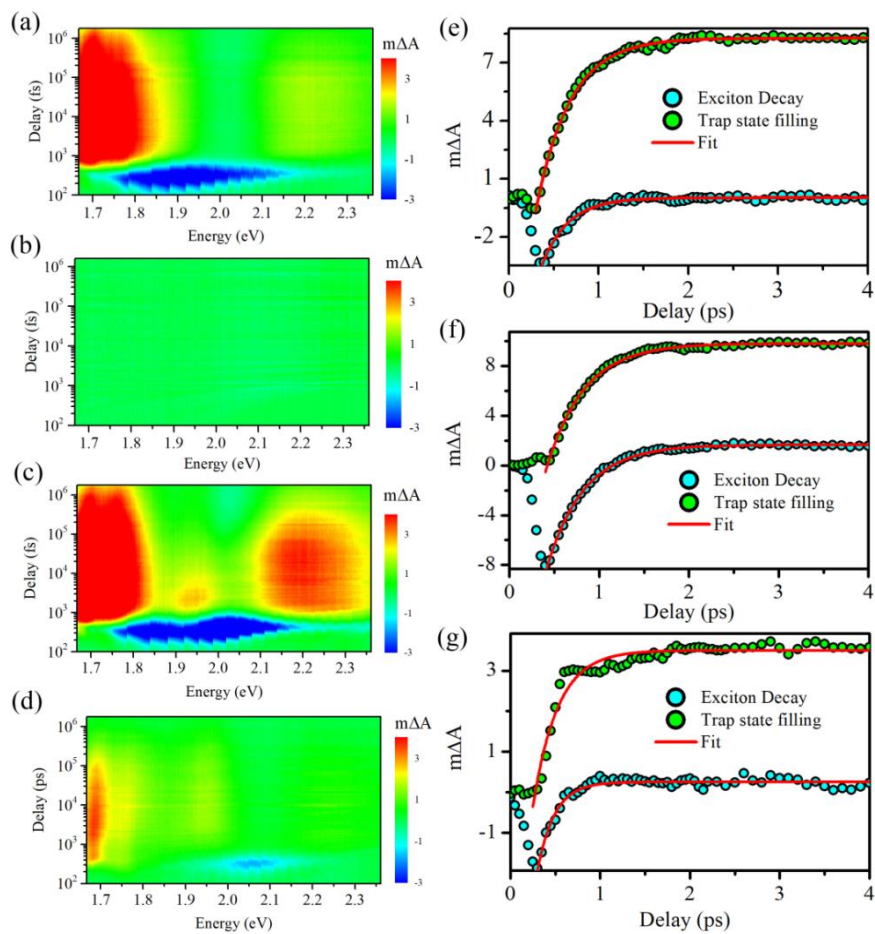

*Figure 8:* Contour plots (see text) of TA measurements for (a) MoS$_2$ (1000), (b) TiO$_2$ (3600) (c) HS1(1000) and (d) HS2(1000). Kinetics for A-exciton decay at 2 eV and trap state filling at 1.73 eV for (e) MoS$_2$(1000), (f) HS1(1000) and (g) HS2(1000).

MoS$_2$ (1000), being an indirect band-gap system, shows a broad exciton band, where the A and B exciton merge ranging from 1.75 eV to 2.21 eV (Fig. 8(a)). The exciton decay is followed by the trap state absorption at ~ 1.69 eV. The exciton decay time (0.29±0.01 fs) is smaller than the trap state build-up time (0.33±0.01 fs). In MoS$_2$, presences of SV induced mid-gap trap states are well known [20, 63]. Whereas, TiO$_2$ is devoid of any exciton band (Fig. 8(b)), for HS1(1000) (Fig 8(c)), well-resolved A and B excitons are observed with larger intensity for B excitons, resembling well with the PL behaviour. Additionally, there is an escalation of the sub-bandgap as well as higher energy delocalized trap states after the decay of excitons, which implies that after the dissociation of excitons, the carriers may be transported from MoS$_2$ to the delocalized acceptor levels of TiO$_2$ near VB. On the other hand, for HS2 (1000) (Fig 8(d)), exciton signals are highly quenched, implying the depletion of carriers from the MoS$_2$ to TiO$_2$, supporting a stronger *p*-type doping behaviour. The presence of localized mid-gap traps is also observed in HS2 (1000), resembling the band-structure of both pristine HS as well as in the presence of OV and SV (Fig 5). Therefore, HS1 manifests a longer lifetime of excitons than HS2, as also seen in the Table 2, which can be imputed to the presence of larger interlayer charge transfer and greater density of mid-gap traps. Fig S9 and S10 show similar behaviour for other HS too. The comparative TA spectra recorded with a delay of 2 ps, shows a similar behaviour, as presented in Fig S8(c). In the next section, we examined the presence of the mid-gap traps, responsible for the quenching of excitons, in different HS after comparing with the pristine system with the help of *z*-scan two-photon absorption (TPA) measurements.

Thus, the interdependence of mid-gap trap states and the manifestation of exciton signals for these two types of HS provides an idea about the presence of vacancies and their respective densities at the interfacial region.

### 3.7. Non-linear response in HS: effective two photon absorption measurements

Nanosecond-resolved open aperture Z-scan measurements are performed on the same systems as in the previous section, *viz*. MoS$_2$ (1000), HS1 (1000) and HS2 (1000), to study the quantitative analysis of the third-order nonlinearity and to analyse the possibility of mid-gap states. In this measurement, the normalized transmittance was recorded as a function of the sample positions.

In this measurement, we purposely concentrate on exciting the system with a sub-band-gap 1064 nm near-infrared excitation and thereby compare the possibilities of the two-photon absorption (TPA) for different HS. Higher intensities of below band-gap excitations indicate higher density of the mid-gap levels, which can be attributed to the presence of vacancies in the respective HS. Fig 9(a) presents the normalized transmittance as a function of *z*-scan positions for $MoS_2$ (1000), $TiO_2$ (3600) and HS1 (1000), with a peak intensity of 15 MW/cm$^2$. For both $MoS_2$ and $TiO_2$, 1064 nm corresponds to a sub-bandgap excitation and Fig 9(a) displays a monotonous increase of the normalized transmittance till the focal point (z = 0), representing a typical feature of the saturable absorption (SA) for both systems[63-64]. However, HS1 (1000) indicates highly reduced SA, as compared to both $TiO_2$ and $MoS_2$, as shown in Fig. 9(a). Such behavior corroborates nicely with the unquenched exciton signals for HS1 (1000), indicating a lower density of mid-gap traps. On the other hand, Fig 9(b), depicting a similar comparison for HS2 (1000), manifests a significant enhancement of SA and thereby indicates the presence of an effective TPA in this system. This enhancement of SA due to the presence of effective TPA indicates the presence of mid-gap states within HS2 (1000), providing intermediate levels for TPA. This behavior goes hand in hand with the TA exciton dynamics, where the exciton signals were highly quenched for HS2(1000) due to the depletion of carriers from $MoS_2$ and their subsequent transfer to $TiO_2$. Thus the presence of mid-gap states, as indicated by the quenching of excitons, is also verified by the Z-scan measurements.

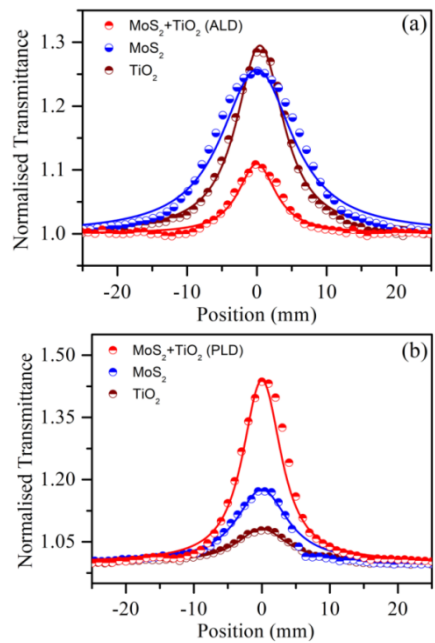

*Figure 9: Comparative normalized transmittance as a function of position in open-aperture Z scan at a peak intensity of 15 MW/cm$^2$ at 7 ns, 1064 nm pulse excitation for (a) HS1(1000), MoS$_2$(1000), TiO$_2$(3600) and (b) HS2(1000), MoS$_2$(1000), TiO$_2$(3600). The solid lines represent theoretical fitting.*

The normalized transmittances for different peak intensities are plotted in Fig 10 for (a) MoS$_2$ (1000), (b) TiO$_2$ (3600), (c) HS1 (1000) and (d) HS2 (1000). For MoS$_2$, TiO$_2$, and HS2, the power dependence is very obvious. However, for HS1 (1000), the plotted transmittances have little dependence on different powers, because of the reduced extent of TPA. A comparison of the Normalized transmittance as a function of input intensity is plotted in Fig 11 for (a) HS1 (1000) and (b) HS2 (1000) at 1064 nm excitation. Optical limiting curve, represented by the variation of the output intensity as a function of input intensity, are plotted in Fig 11 for HS1 (1000) (Fig 11(a) and (b)) and for HS2 (1000) (Fig 11(c) and (d)) respectively. The black line shows the linear transmittance and the curved lines represent the theoretically fitted curve. It is evident from the figure that HS2 is having a more prominent non-linearity than HS1. More thickness of the TiO$_2$ layer in HS2 may be a reason for this increased non-linearity.

Thus, among both the HS, PLD grown HS2 (1000) acts as a better potential saturable absorber material having a prominent non-linear response for the sub-bandgap (1064 nm) excitation.

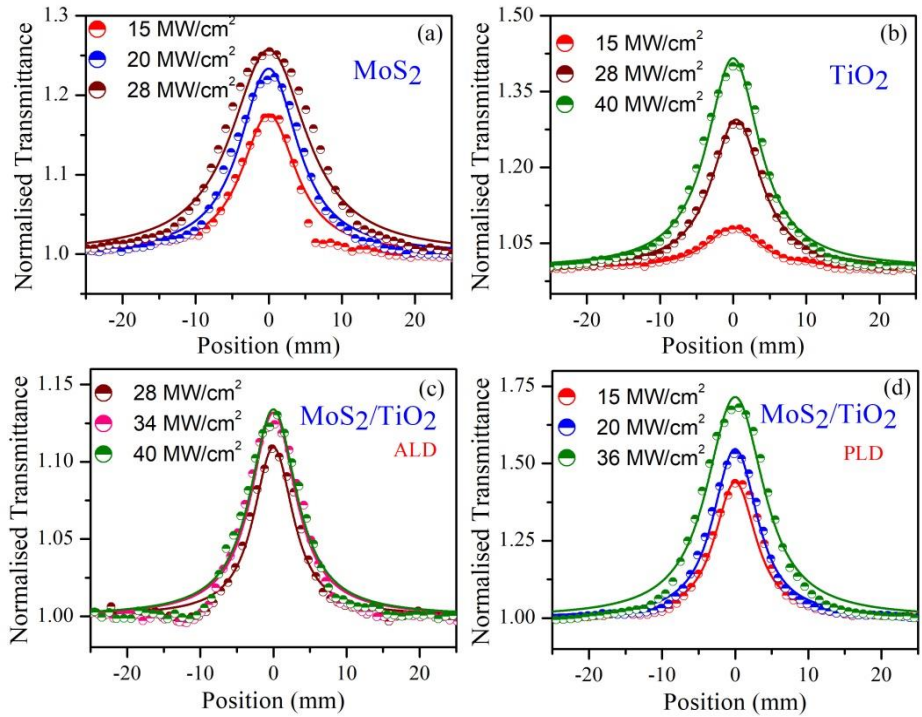

*Figure 10:* *Normalized transmittance as a function of position in open aperture Z scan at 7ns, 1064-nm pulse excitation with different peak intensities for (a) MoS$_2$ (b) TiO$_2$ (c) MoS$_2$/TiO$_2$ (ALD) (HS1) and (d) MoS$_2$/TiO$_2$ (PLD)(HS2) respectively.*

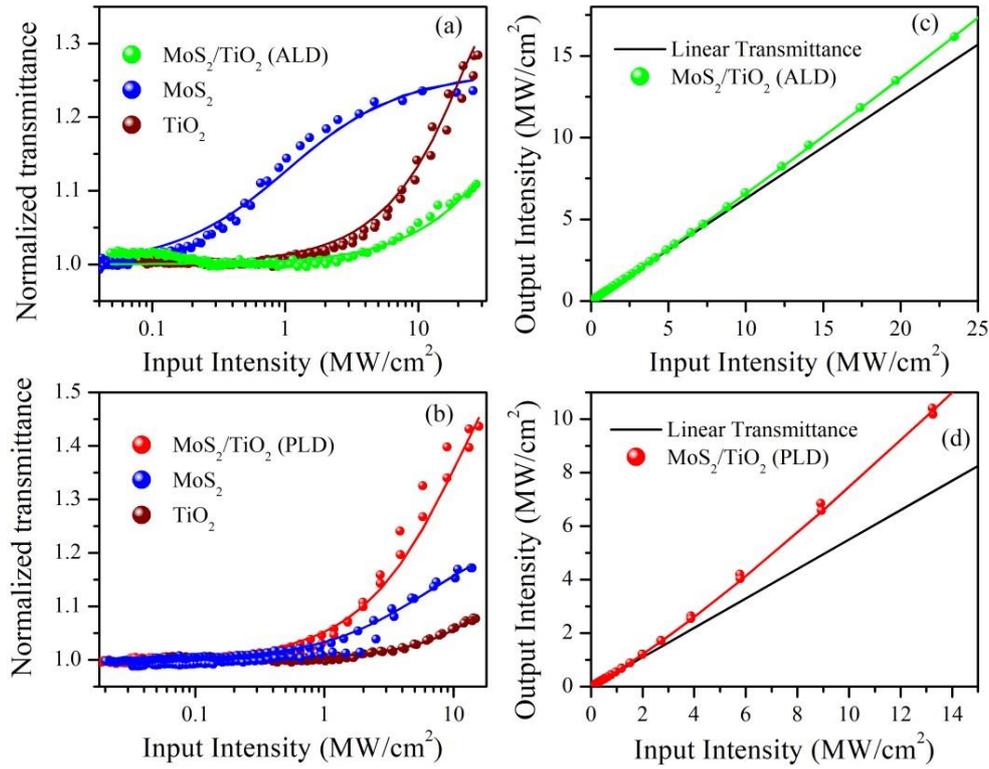

*Figure 11:* Comparison of Normalized transmittance as a function of input intensity for (a) HS1(1000) and (b) HS2(1000) at 1064 nm excitation. Here the symbols and solid curves represent the experimental data and theoretical fit respectively. Optical limiting curve represented by the variation of output intensity as a function of input intensity for (c) HS1(1000) and (d) HS2(1000) respectively. The black line shows the linear transmittance and the curved lines represent the theoretical fit.

### 3.8. Application in four-probe photo-transport and large-area optoelectronics:

The resultant outcome of the previous sections implies a *p*-type doping for both HS1 and HS2 with the extent of doping being more for HS2. To analyse the applicability of such large-area HS for optoelectronic applications, we have measured the four-probe photo-transport and later have analysed its potential for the photo-transistor applications.

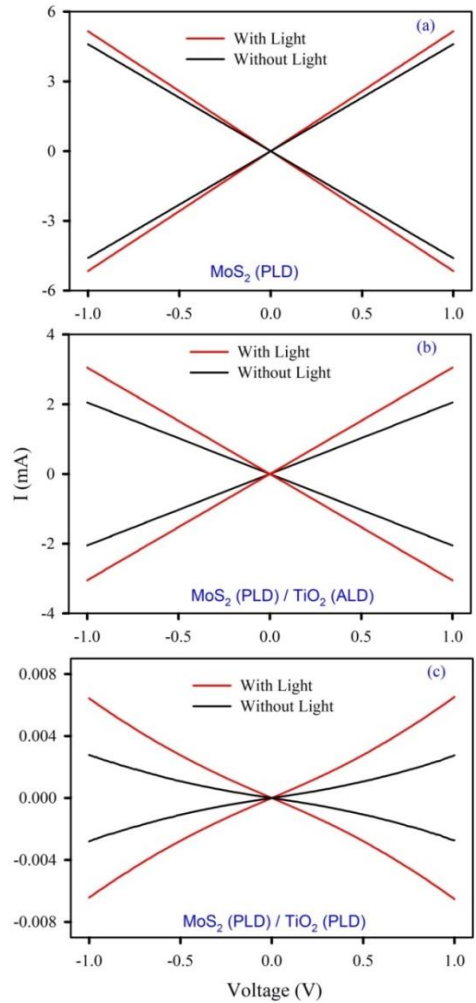

*Figure 12:* Four-probe I-vs-V characteristics of different systems a) MoS$_2$ (1000), b) HS1(1000) and c) HS2(1000).

On MoS$_2$ (1000), HS1 (1000), and HS2 (1000) films, Ni contacts of 500 μm diameter are deposited by thermal evaporation after using a shadow mask array. Earlier reports [65] indicated a betterment of Schottky barrier heights by using Ni-contacts. Fig 12 depicts the results of four-probe photo-transport measurements for these three systems, where channel-width normalized currents (in mA) are plotted as a function of voltages (in volt). The linear current-versus-voltage behaviour confirms the ohmic nature of the Ni-contact for the first two cases. However, for HS2 (1000), with increasing voltage, the characteristics become non-linear. A closer comparison shows that the reduction of current in HS1 (1000) is not so significant in comparison to MoS$_2$ (1000), whereas, there are three orders of magnitude reduction in the case of HS2 (1000). Under illumination, MoS$_2$ (1000) does not demonstrate much improvement of photo-generated carriers, as is expected for multi-layered systems. However, both for HS1 and HS2, there is a significant improvement of the photo-generated carriers, with the latter having a better photo-response. Theoretically, we have established the presence of acceptor level-induced *p*-type doping and a direct band-gap nature for such HS. The present result of improved photoresponse tallies well with the theoretical prediction. As a next step, we have analysed the potential of $MoS_2/TiO_2$ large-area HS in producing large-area phototransistor by fabricating a top-gated device by using the ionic gel as a gate. Since the PLD HS has shown the maximal photoresponse, the device applicability is tested for HS2.

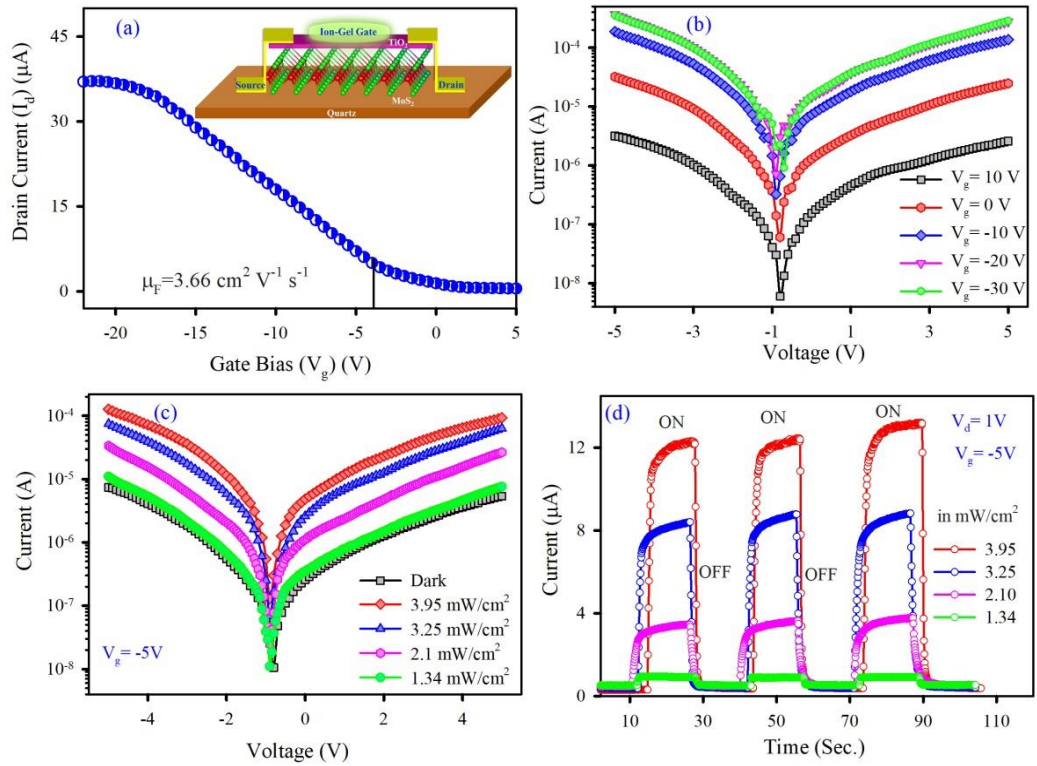

*Figure 13: Phototransistor characteristics of large-area devices from HS. (a) Drain current versus gate voltage characteristics implicating a p-type nature, (b) Drain current versus channel voltage plot for different gate biases, (c) Current versus voltage photo-response under dark and different illuminating condition and (d) Photo-transistor on-off behaviour with different illuminating power.*

Room temperature ionic liquids are well-known [66-68] for their potential application as a dielectric gate for controlling the electronic properties of semiconductors. Usually, field-effect transistors (FET) are designed after using oxide gate dielectrics, offering high operating voltage windows and higher power consumption, rendering them inappropriate for modern-age low-power integrated electronics. The ion-gel dielectric gated FET can overcome these limitations by forming an electric double-layer (EDL) capacitors under an external electric field, due to its large specific capacitance ($> 1\ \mu F.cm^{-2}$)[84].

We demonstrate a low operating voltage EDL-FET based on HS2 (1000) after using ion-gel gate dielectrics and analyse their photo-response characteristics by simultaneous illumination and gate bias. The device fabrication and the measurement details are presented in the Supporting Information. Polyethylene oxide (PEO) and Lithium perchlorate (LiClO$_4$) precursors are used to prepare the gate dielectrics.

Figure 13(a) shows the transfer characteristics ($I_d$ - $V_g$ curves) of the fabricated transistor, measured at a fixed drain-source bias ($V_d$) of 1V, exhibiting a typical FET behaviour. Interestingly, the resulting device exhibits a signature of the *p*-type channel characteristics, predominantly indicating a hole transport through the HS2(1000) channel, which is different from the 2D-MoS$_2$ based *n*-type FET channel behaviour. The off-state current ($I_{off}$) of the FET is ~ 0.53 *μA*, and the channel current increases sharply leading to an ON condition, as $V_g$ decreases below a threshold voltage ($V_{th}$) of -3.5 *V* and then saturates at ~37 *μA* ($I_{on}$) after a gate bias of ~ -18 *V*. The low turn-on voltage (-3.5 *V*) of the fabricated FET, which can be ascribed to the high specific capacitance of the dielectric ion-gel, allows the work function of the channel to be readily modulated at low operating voltages. The resulting on/off ratio ($I_{on}/I_{off}$) are found to be ~75 in between -20 *V* to 5 *V* operating voltages.

Under the application of a negative gate bias ($V_g$) to the channel material (MoS$_2$/TiO$_2$), the resultant electric field drives the Li$^+$ ions towards the gate electrode and the ClO$_4^-$ ions near the channel surface, forming a nanoscale thickness gate capacitor, known as an electric double layer (EDL). The large capacitance of the EDL, leading to a large surface carrier density and induced holes in the *p*-type channel, effectively enhances the electric current through the channel between source and drain electrodes. The field-effect mobility has been extracted by using the following expression:

$$\mu = \frac{\Delta I_d}{\Delta V_g} \times \frac{L}{W C_{sp} V_d}$$

where L is the channel length (1 mm), W is the channel width (2 mm), and $C_{sp}$ is the specific capacitance of the dielectric gel, about ~3 μF/cm$^2$ [68]. From the data presented in Figure 1(a), the field-effect mobility was estimated as ~3.66 $cm^2V^{-1}s^{-1}$. The output characteristics ($I_d$ vs $V_d$) of the ion-gel-gated TiO$_2$–MoS$_2$ heterostructure FET is measured for different fixed gate voltages ($V_g$) and presented in Figure 13(b). The large degree of current modulation in the device, obtained after changing the $V_g$ values, indicates that the field-effect behaviour of the fabricated transistor is dominated by the HS2(1000) channel and not by the contacts.

The current-voltage ($I_d$ – $V_d$) characteristics of the fabricated device for a fixed gate bias (-$V_g$ > $V_{th}$) are shown in Figure 13(c) under dark and various illuminating powers. The symmetric $I$–$V$ nature indicates the formation of a similar junction between Au and HS2(1000) hybrid at both source and drain end. A significant increase of the current level, upon 325 nm monochromatic UV illumination, is due to the photo-generated carriers, leading to the enhancement of the field-effect mobility in the channel. Synergic effect of the dielectric gate and UV illumination, results in substantial enhancement (~10 times) of the dark current for $V_d$ > ±2 V. The photo-switching characteristics of the fabricated EDL-FET photodetector, upon pulsed illumination (with 325 nm laser) with various power densities has been recorded at $V_d$ = 1 $V$ and $V_g$ = -5 $V$ (>$V_{th}$), and the results are presented in Figure 13(d). The rapid and periodic changes in the current level for dark and illuminated conditions represent an excellent reproducibility and stability of the photodetectors. The results also demonstrate that the absence of persistent photo-current and non-fluctuating current level with time. Moreover, the photocurrent response has been increased with increasing illumination intensity. The variation of photocurrent with incident optical intensity has been estimated and observed to display a monotonic increase with incident power, as presented in Fig S11, suggesting that the rate of photo-carrier generation is proportional to the absorbed photon flux. The Responsivity, a key-feature of a photodetector, was calculated for 3.95 mW/cm$^2$ (@325 nm laser wavelength) illumination intensity and is found to be ~3025 A/W, indicating its superiority for millimeter-scale devices.

Thus, in the EDL-FET device, electron-hole pairs are generated in the HS2(1000) channel under illumination and EDL induced positive charges in the channel, which, through the capacitive coupling of the applied negative gate bias, effectively increases the carrier concentration, leading to an enhanced photocurrent. The gate bias is also advantageous to the neutralization of the charged vacancies, which in turn reduces the potential scattering leading to the enhanced field-effect mobility, resulting in an efficient collection of photo-carriers.

Therefore, our results provide an important step towards the realization of the solution-processed, room-temperature deposition of high-capacitance ion-gel dielectrics integrated with 2D-heterostructure based millimeter-scale FETs, for low-power optoelectronic applications.

## 4. Conclusion

This consolidated study demonstrates the electronic and optical impacts of different growth techniques on large area interfaces of $MoS_2$/A-$TiO_2$ HS. Whereas, PLD produces a polycrystalline $TiO_2$ layer with enhanced thickness, ALD results into an amorphous and thinner $TiO_2$ layer. Remarkably, both of these interfaces emanate a *p*-type doping of the $MoS_2$ under-layer with the increasing trend of the extent of doping with the thickness of the $TiO_2$ layer. The resulting *p*-type doping is validated by the first-principles investigation, by demonstrating the impact of the terminating atomic layer of $TiO_2$ on $MoS_2$ and the interfacial vacancies. Substantiating the theoretical results, the steady state and ultrafast optical responses demonstrate that the $MoS_2$/$TiO_2$ (ALD) interface is relatively free of mid-gap localized trap states, having clear signatures of both A and B excitons of $MoS_2$. The $MoS_2$/$TiO_2$ (PLD) interface, on the other hand, has significantly quenched excitons due to a higher density of localized traps. The *z*-scan two-photon absorption of these two interfaces indicates the application of the $MoS_2$(PLD)/$TiO_2$(PLD) interface as a potential saturable absorber having significant non-linear response. The same interface also exhibits a better photoresponse and as an application, we have demonstrated the utilization of this interface in large-area phototransistors having significant hole-mobility and photoresponse. In conclusion, with the help of both theoretical and experimental investigations, we have thoroughly scrutinized the different types of large-area interfaces of $MoS_2$/$TiO_2$ by means of different optical and transport studies and at the end, demonstrate their applications as a potential phototransistor material.

## Acknowledgments

T.K.M. acknowledges the support of DST India for the INSPIRE Research Fellowship and SNBNCBS for funding. D.K. would like to acknowledge the BARC ANUPAM supercomputing facility for computational resources, BRNS CRP on Graphene Analogues for support and motivation, SAIF-IITB for TEM instrument, A. K. Rajarajan of SSPD, BARC for furnace use, S. P. Chakraborty of MSD, BARC for instrumental help of PLD, C L Prajapat for initial attempt of mixed phase deposition by PLD, Madangopal Krishnan for helps in accessing the departmental facilities.

## Supporting Information

See supporting information for details of formation of heterostructure stacks, DFT results of Ti-terminated interface, details of thickness and roughness measurements by AFM, additional data for different HS including XPS, PL, TA and TPA measurements, details of photo-transistor device-fabrication process and tables for thickness and roughness measurements, Raman peak positions, peak shifts of XPS data and time scales for TA.

# References

1. Novoselov, K. S.; Geim, A. K.; Morozov, S. V.; Jiang, D.; Zhang, Y.; Dubonos, S. V.; Grigorieva, I. V.; Firsov, A. A., Electric Field Effect in Atomically Thin Carbon Films. *science* **2004,** *306*, 666-669.
2. Andleeb, S.; Eom, J.; Rauf Naz, N.; Singh, A. K., MoS2 Field-Effect Transistor with Graphene Contacts. *J. Mater. Chem. C* **2017,** *5*, 8308-8314.
3. Kharadi, M. A.; Malik, G. F. A.; Shah, K. A.; Khanday, F. A., Sub-10-nm Silicene Nanoribbon Field Effect Transistor. *IEEE T Electron Dev.* **2019,** *66*, 4976-4981.
4. Bayani, A. H.; Dideban, D.; Vali, M.; Moezi, N., Germanene Nanoribbon Tunneling Field Effect Transistor (GeNR-TFET) with a 10 nm Channel Length: Analog Performance, Doping and Temperature Effects. *Semicond. Sci. Technol.* **2016,** *31*, 045009.
5. Li, L.; Yu, Y.; Ye, G. J.; Ge, Q.; Ou, X.; Wu, H.; Feng, D.; Chen, X. H.; Zhang, Y., Black Phosphorus Field-Effect Transistors. *Nat. Nanotech.* **2014,** *9*, 372-377.
6. Kang, J.; Liu, W.; Sarkar, D.; Jena, D.; Banerjee, K., Computational Study of Metal Contacts to Monolayer Transition-Metal Dichalcogenide Semiconductors. *Phys. Rev. X* **2014,** *4*, 031005.
7. Chhowalla, M.; Shin, H. S.; Eda, G.; Li, L.-J.; Loh, K. P.; Zhang, H., The Chemistry of Two-Dimensional Layered Transition Metal Dichalcogenide Nanosheets. *Nat. Chem.* **2013,** *5*, 263.
8. Fiori, G.; Bonaccorso, F.; Iannaccone, G.; Palacios, T.; Neumaier, D.; Seabaugh, A.; Banerjee, S. K.; Colombo, L., Electronics Based on Two-Dimensional Materials. *Nat. Nanotechnol.* **2014,** *9*, 768-779.
9. Baugher, B. W. H.; Churchill, H. O. H.; Yang, Y.; Jarillo-Herrero, P., Intrinsic Electronic Transport Properties of High-Quality Monolayer and Bilayer MoS2. *Nano Lett.* **2013,** *13*, 4212-4216.
10. Maji, T. K.; Vaibhav, K.; Pal, S. K.; Majumdar, K.; Adarsh, K. V.; Karmakar, D., Intricate Modulation of Interlayer Coupling at The Graphene Oxide/MoSe2 Interface: Application in Time-Dependent Optics and Device Transport. *Phys. Rev. B* **2019,** *99*, 115309
11. Geim, A. K.; Grigorieva, I. V., Van der Waals heterostructures. *Nature* **2013,** *499*, 419-425.
12. Wang, Y.; Kim, J. C.; Wu, R. J.; Martinez, J.; Song, X.; Yang, J.; Zhao, F.; Mkhoyan, A.; Jeong, H. Y.; Chhowalla, M., Van der Waals Contacts between Three-Dimensional Metals and Two-Dimensional Semiconductors. *Nature* **2019,** *568*, 70-74.
13. Kim, C.; Moon, I.; Lee, D.; Choi, M. S.; Ahmed, F.; Nam, S.; Cho, Y.; Shin, H.-J.; Park, S.; Yoo, W. J., Fermi Level Pinning at Electrical Metal Contacts of Monolayer Molybdenum Dichalcogenides. *ACS Nano* **2017,** *11*, 1588-1596.
14. Li, S.-L.; Tsukagoshi, K.; Orgiu, E.; Samorì, P., Charge Transport and Mobility Engineering in Two-Dimensional Transition Metal Chalcogenide Semiconductors. *Chem. Soc. Rev.* **2016,** *45*, 118-151.
15. Das, S.; Chen, H.-Y.; Penumatcha, A. V.; Appenzeller, J., High Performance Multilayer MoS2 Transistors with Scandium Contacts. *Nano Lett.* **2013,** *13*, 100-105.

16. Illarionov, Y. Y.; Rzepa, G.; Waltl, M.; Knobloch, T.; Grill, A.; Furchi, M. M.; Mueller, T.; Grasser, T., The Role of Charge Trapping in MoS2 /SiO2 and MoS2 /hBN Field-Effect Transistors. *2D Mater.* **2016,** *3*, 035004.
17. Dolui, K.; Rungger, I.; Sanvito, S., Origin of the n-type and p-type Conductivity of MoS2 Monolayers on a SiO2 Substrate. *Phys. Rev. B* **2013,** *87*, 165402.
18. Bao, W.; Cai, X.; Kim, D.; Sridhara, K.; Fuhrer, M. S., High Mobility Ambipolar MoS2 Field-Effect Transistors: Substrate and Dielectric Effects. *Appl. Phys. Lett.* **2013,** *102*, 042104.
19. Maji, T. K.; Bagchi, D.; Kar, P.; Karmakar, D.; Pal, S. K., Enhanced Charge Separation through Modulation of Defect-State in Wide Band-gap Semiconductor for Potential Photocatalysis Application: Ultrafast Spectroscopy and Computational Studies. *J Photochem. Photobiol. A: Chem.* **2017,** *332*, 391-398.
20. Karmakar, D.; Halder, R.; Padma, N.; Abraham, G.; Vaibhav, K.; Ghosh, M.; Kaur, M.; Bhattacharya, D.; Chandrasekhar Rao, T., Optimal Electron Irradiation as a Tool for Functionalization of MoS2: Theoretical and Experimental Investigation. *J Appl. Phys.* **2015,** *117*, 135701.
21. Nipane, A.; Karmakar, D.; Kaushik, N.; Karande, S.; Lodha, S., Few-Layer MoS2 p-type Devices Enabled by Selective Doping using Low Energy Phosphorus Implantation. *ACS Nano* **2016,** *10*, 2128-2137.
22. Srinivas, V.; Barik, S. K.; Bodo, B.; Karmakar, D.; Chandrasekhar Rao, T. V., Magnetic and Electrical Properties of Oxygen Stabilized Nickel Nanofibers Prepared by the Borohydride Reduction Method. *J. Magn. Magn. Mater.* **2008,** *320*, 788-795.
23. Kaushik, N.; Karmakar, D.; Nipane, A.; Karande, S.; Lodha, S., Interfacial n-Doping Using an Ultrathin TiO2 Layer for Contact Resistance Reduction in MoS2. *ACS Appl. Mater. Interfaces* **2016,** *8*, 256-263.
24. Cowley, A. M.; Sze, S. M., Surface States and Barrier Height of Metal-Semiconductor Systems. *J Appl. Phys.* **1965,** *36*, 3212-3220.
25. Kim, G.-S.; Kim, S.-H.; Lee, T. I.; Cho, B. J.; Choi, C.; Shin, C.; Shim, J. H.; Kim, J.; Yu, H.-Y., Fermi-Level Unpinning Technique with Excellent Thermal Stability for n-Type Germanium. *ACS Appl. Mater. Interfaces* **2017,** *9*, 35988-35997.
26. Kim, G.-S.; Lee, T. I.; Cho, B. J.; Yu, H.-Y., Schottky Barrier Height Modulation of Metal–Interlayer–Semiconductor Structure Depending on Contact Surface Orientation for Multi-Gate Transistors. *Appl. Phys. Lett.* **2019,** *114*, 012102
27. Agrawal, A.; Lin, J.; Barth, M.; White, R.; Zheng, B.; Chopra, S.; Gupta, S.; Wang, K.; Gelatos, J.; Mohney, S. E.; Datta, S., Fermi Level Depinning and Contact Resistivity Reduction using a Reduced Titania Interlayer in n-Silicon Metal-Insulator-Semiconductor Ohmic Contacts. *Appl. Phys. Lett.* **2014,** *104*, 112101.
28. Kim, G.-S.; Kim, S.-W.; Kim, S.-H.; Park, J.; Seo, Y.; Cho, B. J.; Shin, C.; Shim, J. H.; Yu, H.-Y., Effective Schottky Barrier Height Lowering of Metal/n-Ge with a TiO2/GeO2 Interlayer Stack. *ACS Appl. Mater. Interfaces* **2016,** *8*, 35419-35425.
29. Zhao, Y.; Xu, K.; Pan, F.; Zhou, C.; Zhou, F.; Chai, Y., Doping, Contact and Interface Engineering of Two-Dimensional Layered Transition Metal Dichalcogenides Transistors. *Adv. Functional Mater.* **2017,** *27*, 1603484.

30. Kang, J.; Liu, W.; Banerjee, K., High-Performance MoS2 Transistors with Low-Resistance Molybdenum Contacts. *Appl. Phys. Lett.* **2014,** *104*, 093106.
31. Freedy, K. M.; Olson, D. H.; Hopkins, P. E.; McDonnell, S. J., Titanium Contacts to MoS2 with Interfacial Oxide: Interface Chemistry and Thermal Transport. *Phys. Rev. Mater.* **2019,** *3*, 104001.
32. Yu, Z.; Pan, Y.; Shen, Y.; Wang, Z.; Ong, Z.-Y.; Xu, T.; Xin, R.; Pan, L.; Wang, B.; Sun, L.; Wang, J.; Zhang, G.; Zhang, Y. W.; Shi, Y.; Wang, X., Towards Intrinsic Charge Transport in Monolayer Molybdenum Disulfide by Defect and Interface Engineering. *Nat. Commun.* **2014,** *5*, 5290.
33. Szabó, Á.; Jain, A.; Parzefall, M.; Novotny, L.; Luisier, M., Electron Transport through Metal/MoS2 Interfaces: Edge- or Area-Dependent Process? *Nano Lett.* **2019,** *19*, 3641-3647.
34. Park, W.; Min, J. W.; Shaikh, S. F.; Hussain, M. M., Stable MoS2 Field-Effect Transistors Using TiO2 Interfacial Layer at Metal/MoS2 Contact. *Phys. Status Solidi (A)* **2017,** *214*, 1700534.
35. Neupane, G. P.; Tran, M. D.; Yun, S. J.; Kim, H.; Seo, C.; Lee, J.; Han, G. H.; Sood, A. K.; Kim, J., Simple Chemical Treatment to n-Dope Transition-Metal Dichalcogenides and Enhance the Optical and Electrical Characteristics. *ACS Appl. Mater. Interfaces* **2017,** *9*, 11950-11958.
36. Kufer, D.; Lasanta, T.; Bernechea, M.; Koppens, F. H. L.; Konstantatos, G., Interface Engineering in Hybrid Quantum Dot–2D Phototransistors. *ACS Photonics* **2016,** *3*, 1324-1330.
37. Nan, F.; Li, P.; Li, J.; Cai, T.; Ju, S.; Fang, L., Experimental and Theoretical Evidence of Enhanced Visible Light Photoelectrochemical and Photocatalytic Properties in MoS2/TiO2 Nanohole Arrays. *J Phys. Chem. C* **2018,** *122*, 15055-15062.
38. Pak, Y.; Park, W.; Mitra, S.; Sasikala Devi, A. A.; Loganathan, K.; Kumaresan, Y.; Kim, Y.; Cho, B.; Jung, G. Y.; Hussain, M. M., Enhanced Performance of MoS2 Photodetectors by Inserting an ALD-Processed TiO$_2$ Interlayer. *Small* **2018,** *14*, 1703176.
39. Lin, J.; Zhong, J.; Zhong, S.; Li, H.; Zhang, H.; Chen, W., Modulating Electronic Transport Properties of MoS2 Field Effect Transistor by Surface Overlayers. *Appl. Phys. Lett.* **2013,** *103*, 063109.
40. Xu, K.; Wang, Y.; Zhao, Y.; Chai, Y., Modulation Doping of Transition Metal Dichalcogenide/Oxide Heterostructures. *J. Mater. Chem. C* **2017,** *5*, 376-381.
41. Chuang, S.; Battaglia, C.; Azcatl, A.; McDonnell, S.; Kang, J. S.; Yin, X.; Tosun, M.; Kapadia, R.; Fang, H.; Wallace, R. M.; Javey, A., MoS2 P-type Transistors and Diodes Enabled by High Work Function MoOx Contacts. *Nano Lett.* **2014,** *14*, 1337-1342.
42. Zhou, C.; Zhao, Y.; Raju, S.; Wang, Y.; Lin, Z.; Chan, M.; Chai, Y., Carrier Type Control of WSe2 Field-Effect Transistors by Thickness Modulation and MoO3 Layer Doping. *Adv. Func. Mater.* **2016,** *26*, 4223-4230.
43. Luo, W.; Zhu, M.; Peng, G.; Zheng, X.; Miao, F.; Bai, S.; Zhang, X.-A.; Qin, S., Carrier Modulation of Ambipolar Few-Layer MoTe2 Transistors by MgO Surface Charge Transfer Doping. *Adv. Func. Mater.* **2018,** *28*, 1704539.

44. Ho, P.-H.; Chang, Y.-R.; Chu, Y.-C.; Li, M.-K.; Tsai, C.-A.; Wang, W.-H.; Ho, C.-H.; Chen, C.-W.; Chiu, P.-W., High-Mobility InSe Transistors: The Role of Surface Oxides. *ACS Nano* **2017,** *11*, 7362-7370.

45. Li, M.; Lin, C.-Y.; Yang, S.-H.; Chang, Y.-M.; Chang, J.-K.; Yang, F.-S.; Zhong, C.; Jian, W.-B.; Lien, C.-H.; Ho, C.-H.; Liu, H.-J.; Huang, R.; Li, W.; Lin, Y.-F.; Chu, J., High Mobilities in Layered InSe Transistors with Indium-Encapsulation-Induced Surface Charge Doping. *Adv. Mater.* **2018,** *30*, 1803690.

46. Ghiasi, T. S.; Quereda, J.; van Wees, B. J., Bilayer h-BN Barriers for Tunneling Contacts in Fully-Encapsulated Monolayer MoSe 2 Field-Effect Transistors. *2D Mater.* **2018,** *6*, 015002.

47. Hattori, Y.; Taniguchi, T.; Watanabe, K.; Nagashio, K., Determination of Carrier Polarity in Fowler–Nordheim Tunneling and Evidence of Fermi Level Pinning at the Hexagonal Boron Nitride/Metal Interface. *ACS Appl. Mater. Interfaces* **2018,** *10*, 11732-11738.

48. Wei, T.; Lau, W. M.; An, X.; Yu, X., Interfacial Charge Transfer in $MoS_2$/$TiO_2$ Heterostructured Photocatalysts: The Impact of Crystal Facets and Defects. *Molecules* **2019,** *24*, 1769

49. Chen, B.; Meng, Y.; Sha, J.; Zhong, C.; Hu, W.; Zhao, N., Preparation of $MoS_2$/$TiO_2$ Based Nanocomposites for Photocatalysis and Rechargeable Batteries: Progress, Challenges, and Perspective. *Nanoscale* **2018,** *10*, 34-68.

50. Kresse, G.; Furthmüller, J., Efficient Iterative Schemes for Ab initio Total-Energy Calculations using a Plane-wave Basis Set. *Phys. Rev. B* **1996,** *54*, 11169-11186.

51. Grimme, S., Semiempirical GGA-type Density Functional Constructed with a Long-Range Dispersion Correction. *J. Comput. Chem.* **2006,** *27*, 1787-1799.

52. Li, H.; Zhang, Q.; Yap, C. C. R.; Tay, B. K.; Edwin, T. H. T.; Olivier, A.; Baillargeat, D., From Bulk to Monolayer $MoS_2$: Evolution of Raman Scattering. *Adv. Func. Mater.* **2012,** *22*, 1385-1390.

53. Lee, C.; Yan, H.; Brus, L. E.; Heinz, T. F.; Hone, J.; Ryu, S., Anomalous Lattice Vibrations of Single- and Few-Layer $MoS_2$. *ACS Nano* **2010,** *4*, 2695-2700.

54. Zhang, Q.; Ma, L.; Shao, M.; Huang, J.; Ding, M.; Deng, X.; Wei, X.; Xu, X., Anodic Oxidation Synthesis of One-Dimensional $TiO_2$ Nanostructures for Photocatalytic and Field Emission Properties. *J. Nanomater.* **2014,** *2014*, 1-14.

55. Yang, L.; Majumdar, K.; Liu, H.; Du, Y.; Wu, H.; Hatzistergos, M.; Hung, P. Y.; Tieckelmann, R.; Tsai, W.; Hobbs, C.; Ye, P. D., Chloride Molecular Doping Technique on 2D Materials: $WS_2$ and $MoS_2$. *Nano Lett.* **2014,** *14*, 6275-6280.

56. Lin, J. D.; Han, C.; Wang, F.; Wang, R.; Xiang, D.; Qin, S.; Zhang, X.-A.; Wang, L.; Zhang, H.; Wee, A. T. S.; Chen, W., Electron-Doping-Enhanced Trion Formation in Monolayer Molybdenum Disulfide Functionalized with Cesium Carbonate. *ACS Nano* **2014,** *8*, 5323-5329.

57. Smidstrup, S.; Markussen, T.; Vancraeyveld, P.; Wellendorff, J.; Schneider, J.; Gunst, T.; Verstichel, B.; Stradi, D.; Khomyakov, P. A.; Vej-Hansen, U. G.; Lee, M.-E.; Chill, S. T.; Rasmussen, F.; Penazzi, G.; Corsetti, F.; Ojanperä, A.; Jensen, K.; Palsgaard, M. L. N.; Martinez, U.; Blom, A.; Brandbyge, M.; Stokbro, K., QuantumATK: An Integrated Platform of Electronic and Atomic-Scale Modelling Tools. *J Phys. Condens. Matter* **2019,** *32*, 015901.

58.	Smidstrup, S.; Stradi, D.; Wellendorff, J.; Khomyakov, P. A.; Vej-Hansen, U. G.; Lee, M.-E.; Ghosh, T.; Jónsson, E.; Jónsson, H.; Stokbro, K., First-Principles Green's-Function Method for Surface Calculations: A Pseudopotential Localized Basis Set Approach. *Phys. Rev. B* **2017,** *96*, 195309.

59.	Korn, T.; Heydrich, S.; Hirmer, M.; Schmutzler, J.; Schüller, C., Low-Temperature Photocarrier Dynamics in Monolayer MoS2. *Appl. Phys. Lett.* **2011,** *99*, 102109.

60.	Elsaesser, T.; Shah, J.; Rota, L.; Lugli, P., Initial Thermalization of Photoexcited Carriers in GaAs Studied by Femtosecond Luminescence Spectroscopy. *Phys. Rev. Lett.* **1991,** *66*, 1757-1760.

61.	Das, S.; Gupta, G.; Majumdar, K., Layer Degree of Freedom for Excitons in Transition Metal Dichalcogenides. *Phys. Rev. B* **2019,** *99*, 165411.

62.	Sim, S.; Park, J.; Song, J.-G.; In, C.; Lee, Y.-S.; Kim, H.; Choi, H., Exciton Dynamics in Atomically Thin MoS2: Interexcitonic Interaction and Broadening Kinetics. *Phys. Rev. B* **2013,** *88*, 075434.

63.	Sharma, R.; Aneesh, J.; Yadav, R. K.; Sanda, S.; Barik, A. R.; Mishra, A. K.; Maji, T. K.; Karmakar, D.; Adarsh, K. V., Strong Interlayer Coupling Mediated Giant Two-Photon Absorption in MoSe2/Graphene Oxide Heterostructure: Quenching of Exciton Bands. *Phys. Rev. B* **2016,** *93*, 155433.

64.	Yadav, R. K.; Aneesh, J.; Sharma, R.; Abhiramnath, P.; Maji, T. K.; Omar, G. J.; Mishra, A. K.; Karmakar, D.; Adarsh, K. V., Designing Hybrids of Graphene Oxide and Gold Nanoparticles for Nonlinear Optical Response. *Phys. Rev. Appl.* **2018,** *9*, 044043.

65.	Gupta, G.; Kallatt, S.; Majumdar, K., Direct Observation of Giant Binding Energy Modulation of Exciton Complexes in Monolayer MoSe2. *Phys. Rev. B* **2017,** *96*, 081403(R).

66.	Clement, C. E.; Jiang, D.; Thio, S. K.; Park, S.-Y., A Study of Dip-Coatable, High-Capacitance Ion Gel Dielectrics for 3D EWOD Device Fabrication. *Materials* **2017,** *10*, 41.

67.	Taghavikish, M.; Subianto, S.; Gu, Y.; Sun, X.; Zhao, X. S.; Choudhury, N. R., A Poly(ionic liquid) Gel Electrolyte for Efficient all Solid Electrochemical Double-Layer Capacitor. *Sci. Rep.* **2018,** *8*, 10918.

68.	Mondal, S.; Ram Ghimire, R.; Raychaudhuri, A., Enhancing Photoresponse by Synergy of Gate and Illumination in Electric Double Layer Field Effect Transistors Fabricated on n-ZnO. *Appl. Phys. Lett.* **2013,** *103*, 231105.

*Supporting Information for*

# Combinatorial Large-area MoS$_2$/Anatase-TiO$_2$ interface: A Pathway to Emergent Optical and Opto-electronic Functionalities

*Tuhin Kumar Maji [a], J R Aswin [b], Subhrajit Mukherjee [c], Rajath Alexander [d], Anirban Mondal [b], Sarthak Das [e], R. K. Sharma [f], N. K. Chakraborty [g], K. Dasgupta [d], Anjanashree M R Sharma [h], Ranjit Hawalder [i], Manjiri Pandey [j], Akshay Naik [h], Kausik Majumdar [e], Samir Kumar Pal [a], K V Adarsh [b], Samit Kumar Ray [a, c], Debjani Karmakar [k, \*]*

[a] Department of Chemical Biological and Macromolecular Sciences, S.N.Bose National Centre for Basic Sciences, Sector III, JD Block, Kolkata 700106, India.
[b] Department of Physics, Indian Institute of Science Education and Research, Bhopal 462066, India.
[c] Department of Physics, IIT Kharagpur, Kharagpur, West Bengal 721302, India.
[d] Advanced Carbon Materials Section, Bhabha Atomic Research Centre, Trombay, Mumbai 400085, India.
[e] Department of Electrical Communication Engineering, Indian Institute of Science, Bangalore 560012, India.
[f] Raja Rammana Centre for Advance Technology, Parmanu Nagar, Sahkar Nagar Extension, 1, CAT Rd, Rajendra Nagar, Indore, Madhya Pradesh 45201.
[g] Material Science Division, Bhabha Atomic Research Centre, Trombay, Mumbai 400085, India.
[h] Centre for Nano Science and Engineering, Indian Institute of Science, Bangalore, Karnataka, 560012.
[i] Centre for Materials for Electronics Technology, Off Pashan Road, Panchwati, Pune- 411008, India.
[j] Accelerator Control Division, Bhabha Atomic Research Centre, Trombay, Mumbai 400085, India.
[k] Technical Physics Division, Bhabha Atomic Research Centre, Trombay 400085, India.

*\*Corresponding Author:* **Dr. Debjani Karmakar**

**E-mail:** *debjan@barc.gov.in*

## AFM of Heterostructure

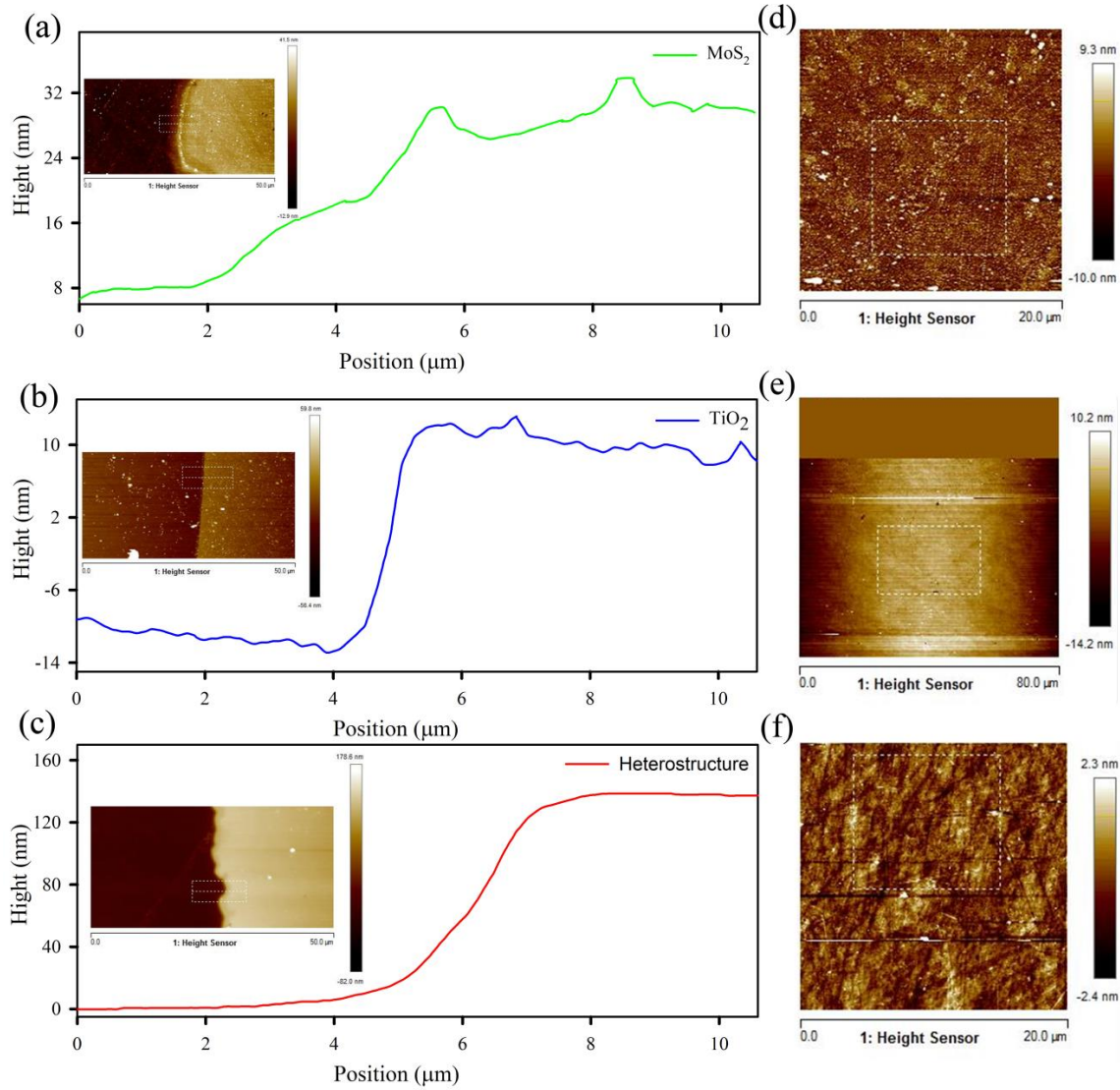

***Figure S1:*** *Thickness of different deposited systems a) MoS$_2$- 1000 Pulse, b) TiO$_2$ 3600 Pulse, c) MoS$_2$ (1000P)/TiO$_2$ (3600P). Surface roughness measurement of a) MoS$_2$- 1000 Pulse, b) TiO$_2$ 3600 Pulse and c) MoS$_2$ (1000P)/TiO$_2$ (3600P).*

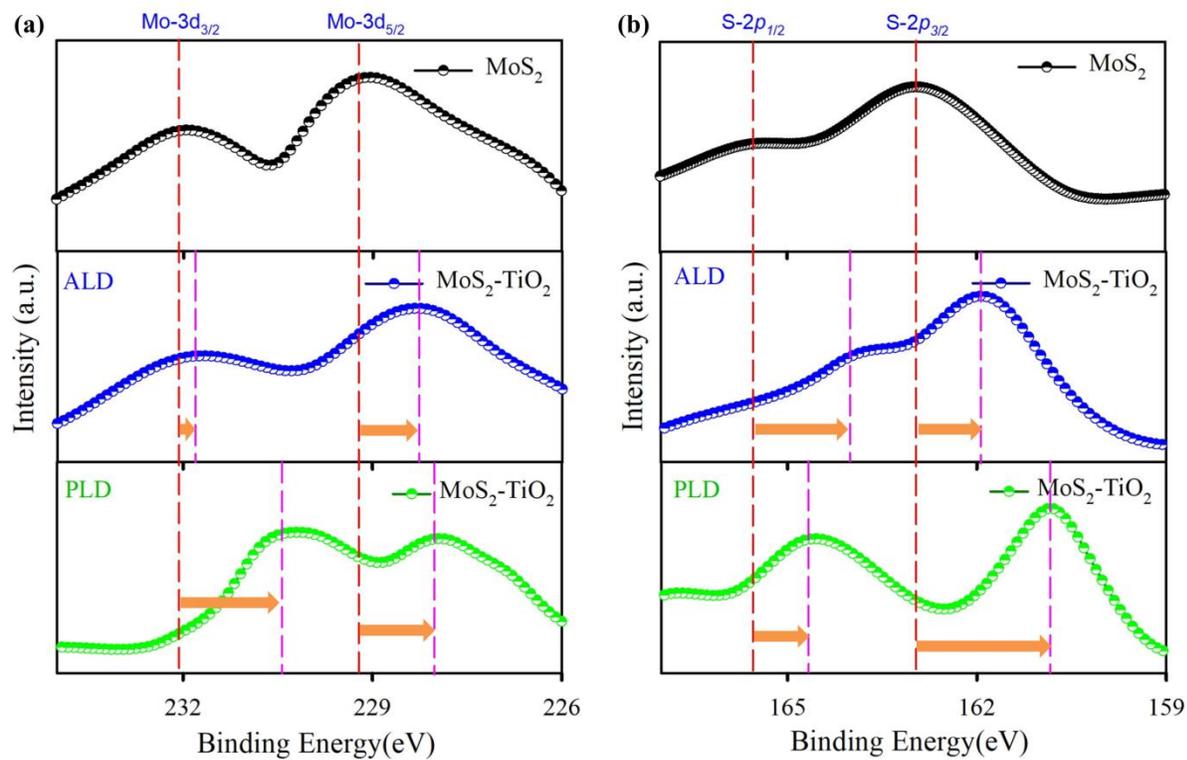

***Figure S2:*** *XPS spectra of MoS$_2$(400P)-TiO$_2$ systems: a) Mo-3d and b) S-2p.*

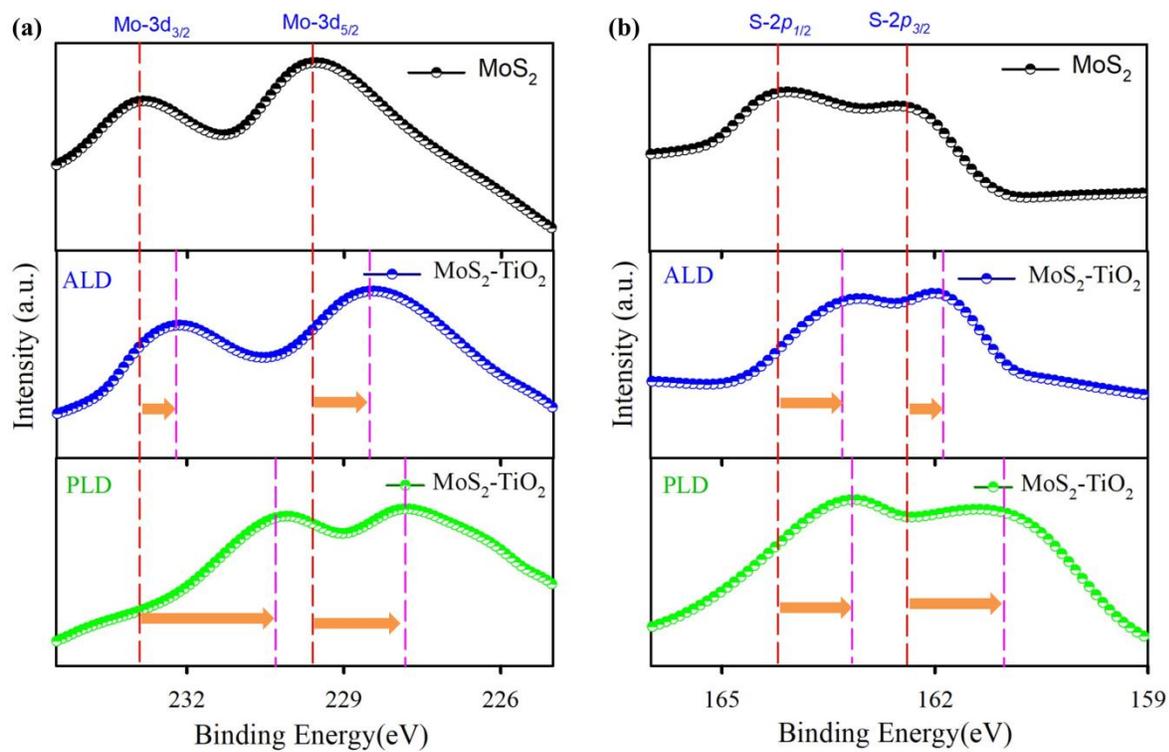

***Figure S3:*** *XPS spectra of MoS$_2$(600P)-TiO$_2$ systems: a) Mo-3d and b) S-2p.*

**Formation of MoS$_2$/TiO$_2$ HS by minimization of interfacial strain:**

The construction of the MoS$_2$/TiO$_2$ interface is done by vertical stacking of a (2×2×1) supercell of 2H-MoS$_2$ (*P6$^3$/mmc*) with a (2×2×1) supercell of anatase TiO$_2$ (*I4$^1$/amd*) by the coincidence site lattice (CSL) method, as implemented in the ATOMISTIC TOOLKIT 15.1 package[1-2]. For minimization of the interfacial mutual rotational strain, a survey was performed through the grid *m***v$_1$** + *n***v$_2$**, with the vectors **v$_1$** and **v$_2$** being the basis vectors of MoS$_2$, so that for the maximum value of the integers *m* and *n*, the supercell of both of the lattices has the lowest mismatch. The surface cells of MoS$_2$ and TiO$_2$ are depicted in Fig S4 (a) and (b) respectively. Next, the mutual strain is minimized by varying the mutual rotation angle between the MoS$_2$ and TiO$_2$ surfaces around the stacking direction (*c*-axis for the present case) in increments of one degree. In all of our calculations, the minimized mutual strain is ~0.87%, when the mutual rotational angle between the two surfaces is ~ 46 degrees. The directional minimal strains $\varepsilon_{11}$, $\varepsilon_{22}$ and $\varepsilon_{33}$ are calculated to be 0.37%, 0.30%, and 1.93% respectively, while the mutual rotational angle between the two surfaces is ~ 46 degrees. The mean absolute strain is calculated to be 0.87%.

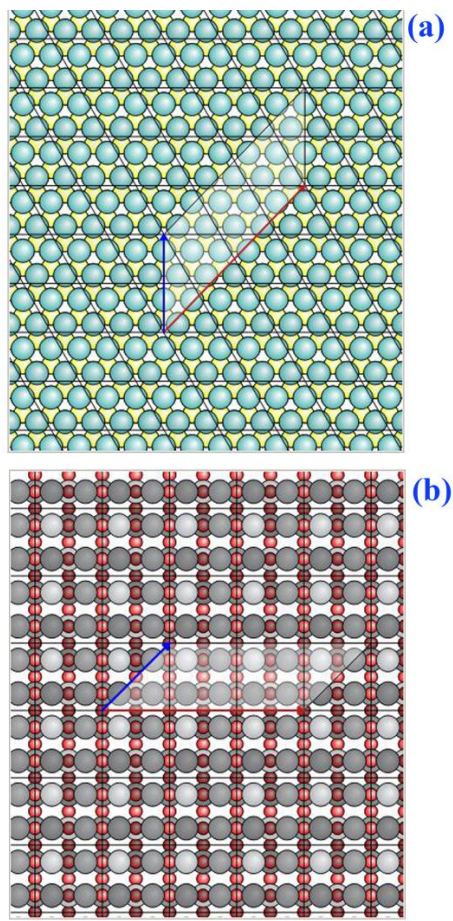

***Figure S4:*** *Selection of the grid and the corresponding two-dimensional lattice vectors for a) MoS$_2$ and b) TiO$_2$ for the construction of hetero-structures.*

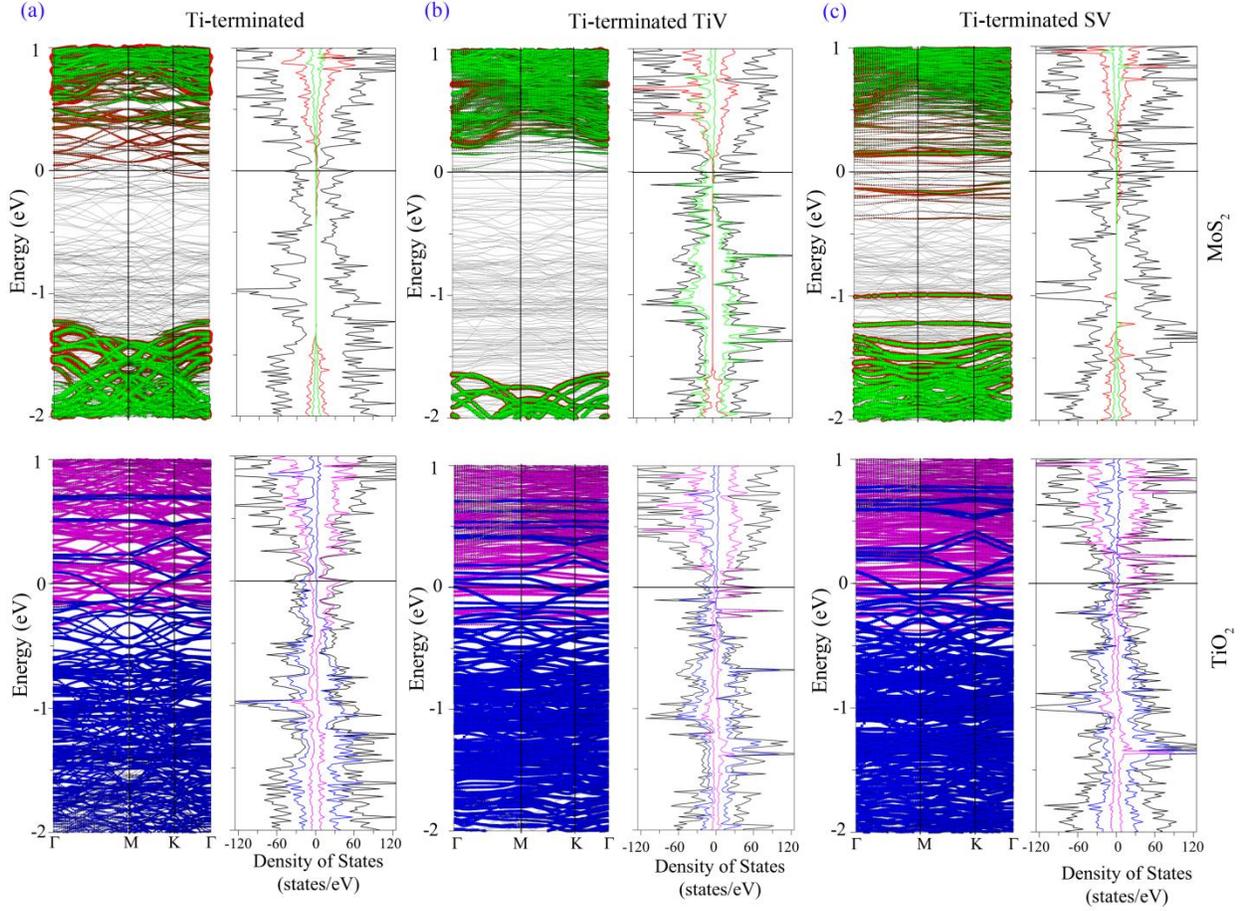

*Figure S5:* Layer and orbital projected fatband and corresponding partial DOS for a) Ti-terminated $MoS_2$-$TiO_2$ systems, b) Ti-terminated $MoS_2$-$TiO_2$ systems with Ti-vacancy (TiV) and b) Ti-terminated $MoS_2$-$TiO_2$ systems with S-vacancy(SV). The upper and lower panels designate the $MoS_2$ and $TiO_2$ part of layer projection with the colour code Mo-d(red), S-p(green), Ti-d(magenta), and O-p(blue).

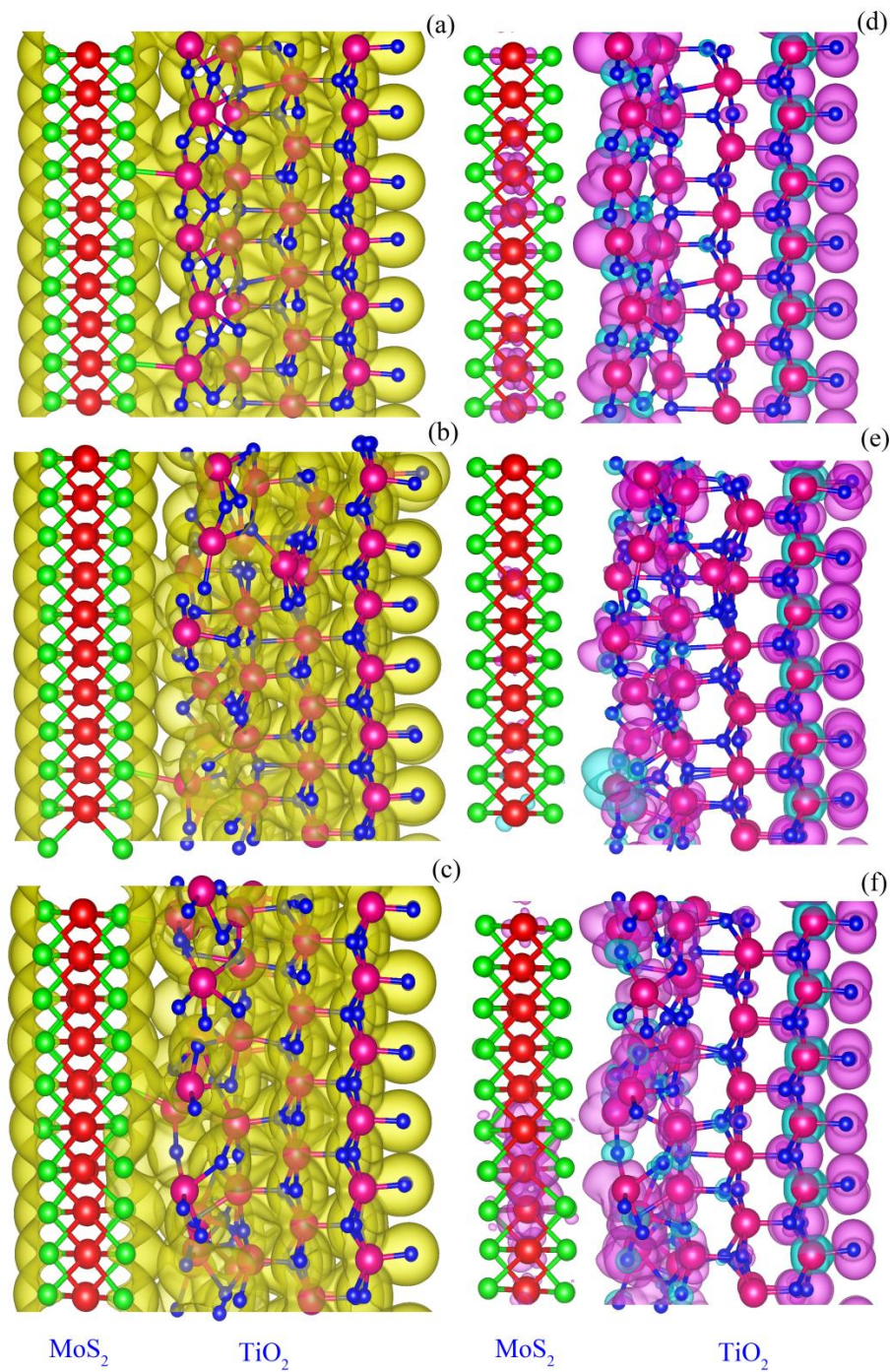

*Figure S6:* *Charge Density of different systems a) Ti-terminated MoS$_2$/TiO$_2$, b) Ti-terminated MoS$_2$/TiO$_2$ with Ti vacancy, c) Ti-terminated MoS$_2$/TiO$_2$ with S vacancy systems and corresponding Spin Density of different systems d) Ti-terminated MoS$_2$/TiO$_2$, e) Ti-terminated MoS$_2$/TiO$_2$ with Ti vacancy, f) Ti-terminated MoS$_2$/TiO$_2$ with S vacancy systems.*

**Steady State PL:**

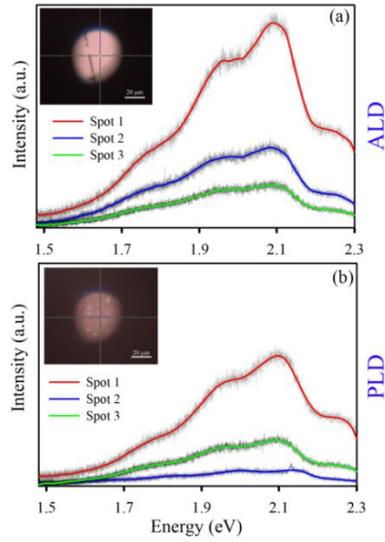

*Figure S7:* *Photoluminescence spectra of a) MoS$_2$ (PLD) - TiO$_2$ (ALD) at 4 K at 3 different random places (inset shows a typical image of a spot) and b) MoS$_2$ (PLD) - TiO$_2$ (PLD) at 4 K at 3 different random places (inset shows a typical image of a spot).*

**Steady State Absorbance**

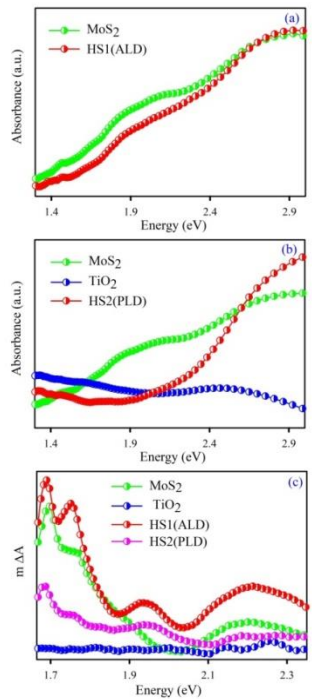

*Figure S8:* *Steady State Absorbance of a) HS1 (ALD), b) HS2 (PLD) and c) Comparison of Transient Absorption of hetero-structures after 2ps delay.*

**Ultrafast Exciton dynamics:**

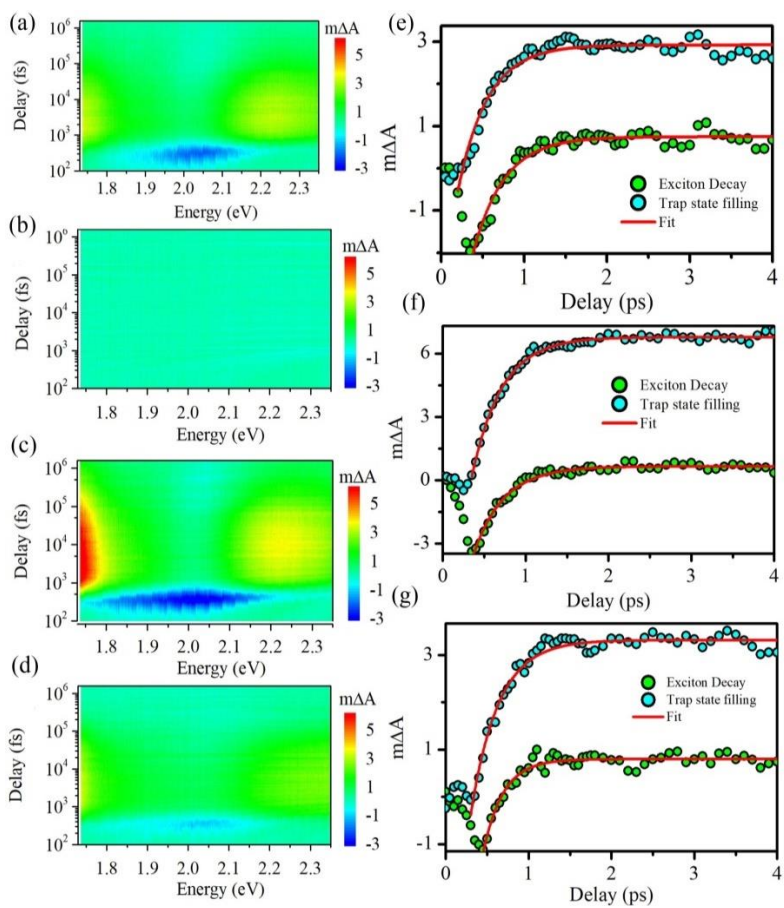

***Figure S9:*** *Contour plot of TA measurements for (a) $MoS_2$ (PLD 400P), (b) $TiO_2$ (PLD 3600P) (c) $MoS_2$ (PLD)/$TiO_2$ (ALD) (d) $MoS_2$ (PLD)/$TiO_2$ (PLD). Kinetics for A-exciton decay at 2 eV and trap state filling at 1.73 eV for (e) $MoS_2$ (PLD 400P), (f) $MoS_2$ (PLD)/$TiO_2$ (ALD) (g) $MoS_2$ (PLD)/$TiO_2$ (PLD).*

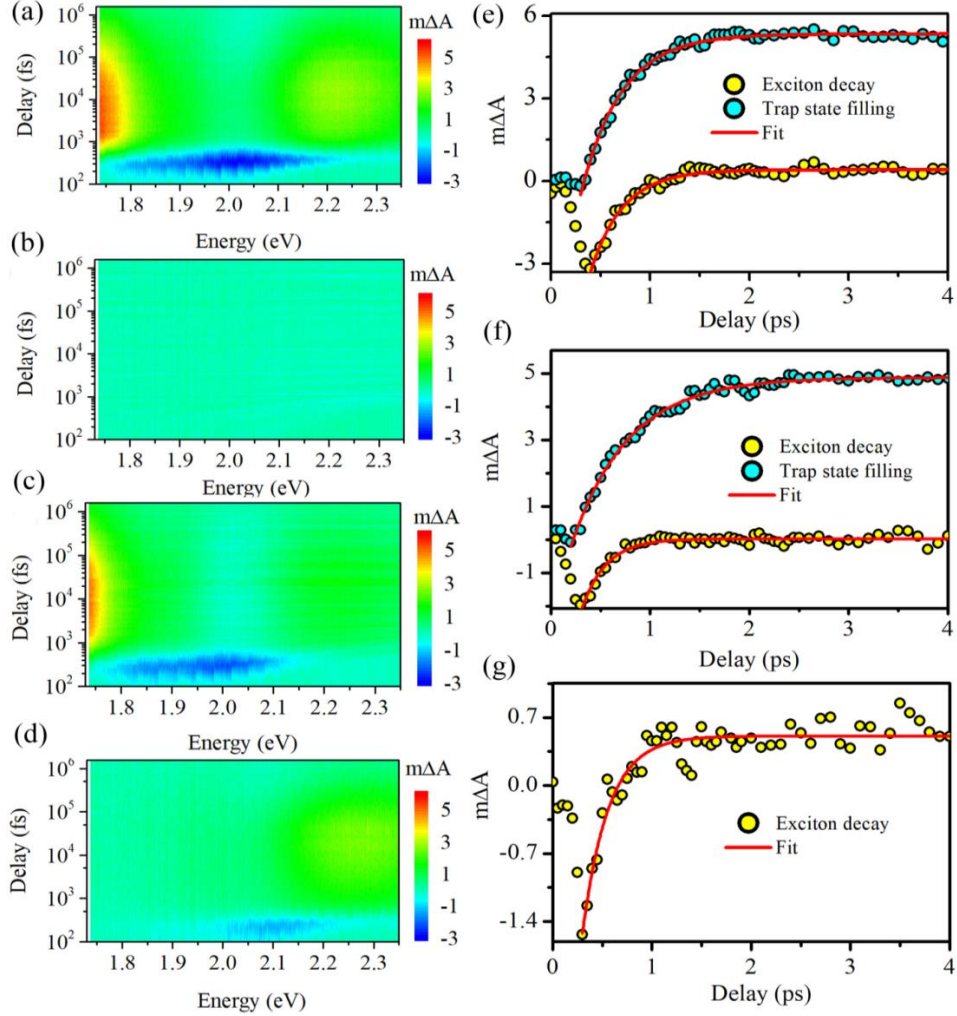

***Figure S10:*** *Contour plot of TA measurements for (a) MoS$_2$ (PLD 600P), (b) TiO$_2$ (PLD 3600P) (c) MoS$_2$ (PLD)/TiO$_2$ (ALD) (d) MoS$_2$ (PLD)/TiO$_2$ (PLD). Kinetics for A-exciton decay at 2 eV and trap state filling at 1.73 eV for (e) MoS$_2$ (PLD 600P), (f) MoS$_2$ (PLD)/TiO$_2$ (ALD) (g) MoS$_2$ (PLD)/TiO$_2$ (PLD).*

**Device fabrication and measurement:**

Thermally evaporated 80 nm thick gold (Au) contact pads were defined as the source and drain electrodes. The device channel is defined by the separation (1 mm) and overlapping (2 mm) length between the two-metal electrodes. A separate coplanar electrode was used as a gate that applies an electric field to the channel through the ion gel. Polyethylene oxide (PEO) and lithium perchlorate (LiClO$_4$) precursors were thoroughly mixed in an 8:1 weight ratio, to form the ion-gel, where the electrolyte is dispersed into the polymeric matrix. The as-prepared gel dielectric was drop-cast such a way that covers the device channel and a portion of the coplanar gate electrode that is positioned remotely from the channel. All the electrical and photo-response measurements were performed in an ambient atmosphere. The output and transfer characteristics of the device are recorded using a Keithley 4200-SCS semiconductor parameter analyzer unit.

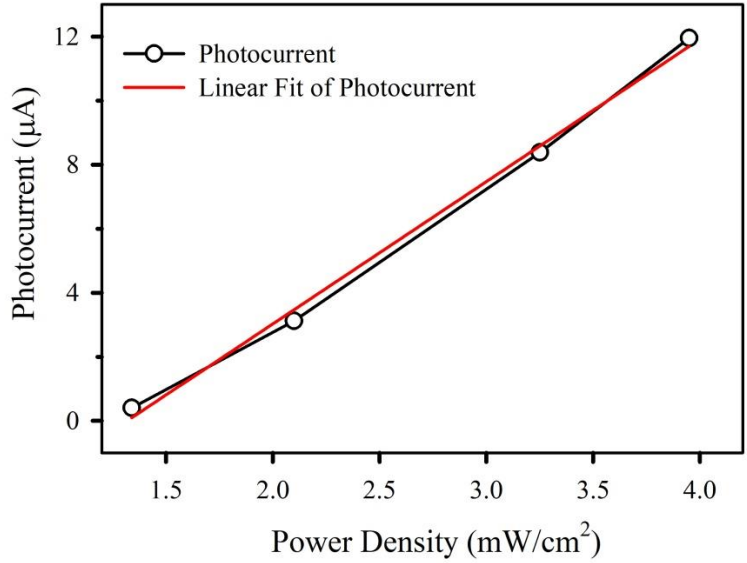

*Figure S11:* *Linear dependence of photocurrent and power density.*

*Table S1:* *Thickness measurement and surface roughness table*

| System | Thickness (nm) | Surface Roughness (nm) |
|---|---|---|
| $MoS_2$-1000P | 28.2 | 2.61 |
| $TiO_2$-3600P | 20.8 | 1.18 |
| Hetero-structure | 136 | 0.86 |

*Table S2:* *Raman Peak of different systems.*

| System | Peak | $MoS_2$ | HS1 | HS2 |
|---|---|---|---|---|
| $MoS_2$ 1000P | $E^1_{2g}$ | 379.1 | 384.8 | 376.9 |
| | $A_{1g}$ | 405.6 | 410.6 | 405.4 |
| $MoS_2$ 600P | $E^1_{2g}$ | 383.3 | 384.6 | 382.3 |
| | $A_{1g}$ | 408.9 | 410.5 | 409.3 |
| $MoS_2$ 400P | $E^1_{2g}$ | 383.5 | 384.6 | 383.7 |
| | $A_{1g}$ | 408.3 | 409.3 | 410.6 |

*Table S3: Peak shift of different systems in XPS.*

| System | | Peak Shift | | | |
|---|---|---|---|---|---|
| | | *Mo 3d$_{3/2}$* | *Mo 3d$_{5/2}$* | *S 2p$_{1/2}$* | *S 2p$_{3/2}$* |
| **MoS$_2$ 1000P** | HS1 | 0.51 | 0.72 | 0.44 | 0.68 |
| | HS2 | 2.18 | 1.07 | 0.64 | 0.95 |
| **MoS$_2$ 600P** | HS1 | 0.69 | 0.98 | 0.86 | 0.46 |
| | HS2 | 2.65 | 1.7 | 1.01 | 1.33 |
| **MoS$_2$ 400P** | HS1 | 0.22 | 0.95 | 1.48 | 1.03 |
| | HS2 | 1.59 | 1.18 | 0.87 | 2.11 |

*Table S4: Time scale for A exciton decay of the different systems (with MoS$_2$ 600 Pulse)*

| Sample | Exciton decay | Trap build up |
|---|---|---|
| MoS$_2$-600P | 0.32±0.01 ps | 0.40±0.01 ps |
| HS1 | 0.34±0.02 ps | 0.39±0.02 ps |
| HS2 | 0.26±0.02 ps | 0.28±0.02 ps |

*Table S5: Time scale for A exciton decay of the different systems (with MoS$_2$ 400 Pulse)*

| Sample | Exciton decay | Trap build up |
|---|---|---|
| MoS$_2$-400P | 0.37±0.02 ps | 0.36±0.02 ps |
| HS1 | 0.36±0.02 ps | 0.36±0.01 ps |
| HS2 | 0.24±0.02 ps | 0.27±0.01 ps |

# References

1. Smidstrup, S.; Markussen, T.; Vancraeyveld, P.; Wellendorff, J.; Schneider, J.; Gunst, T.; Verstichel, B.; Stradi, D.; Khomyakov, P. A.; Vej-Hansen, U. G.; Lee, M.-E.; Chill, S. T.; Rasmussen, F.; Penazzi, G.; Corsetti, F.; Ojanperä, A.; Jensen, K.; Palsgaard, M. L. N.; Martinez, U.; Blom, A.; Brandbyge, M.; Stokbro, K., QuantumATK: An Integrated Platform of Electronic and Atomic-Scale Modelling Tools. *J Phys. Condens. Matter* **2019,** *32*, 015901.

2. Smidstrup, S.; Stradi, D.; Wellendorff, J.; Khomyakov, P. A.; Vej-Hansen, U. G.; Lee, M.-E.; Ghosh, T.; Jónsson, E.; Jónsson, H.; Stokbro, K., First-Principles Green's-Function Method for Surface Calculations: A Pseudopotential Localized Basis Set Approach. *Phys. Rev. B* **2017,** *96*, 195309.